\documentclass[%
amsmath,aps,showkeys,
amssymb,onecolumn,pra,
superscriptaddress,
sort&compress,
preprint,
]{revtex4-2}

\usepackage{url}
\usepackage{hyperref}
\hypersetup{
    colorlinks,
    linkcolor={black},
    citecolor={blue!80!black},
    urlcolor={blue!80!black},
    breaklinks=true
}

\usepackage{graphicx}
\usepackage{bm}
\usepackage{bbold}
\usepackage{dcolumn}
\newcolumntype{d}[1]{D{.}{.}{#1}}

\usepackage[dvipsnames]{xcolor}

\renewcommand{\L}{\langle}

\usepackage{url}
\usepackage{tikz}
\usepackage{pgfplots}
\pgfplotsset{compat=1.14}
\usepackage{booktabs}
\usepackage{pgfplotstable}
\usepackage{float}
\usepackage{subcaption}
\renewcommand{\arraystretch}{1.4} 
\usepackage{multirow}
\usepackage{bbold}
\usepackage{listings}

\usepackage{enumerate}
\usepackage{latexsym}
\usepackage{amsfonts, amsmath, bbm, amssymb, amsthm}
\usepackage{epic}
\usepackage{float}
\usepackage{stfloats}
\usepackage{placeins}

\makeatletter
\g@addto@macro{\UrlBreaks}{\do\/\do\-}
\makeatother

\renewcommand{\arraystretch}{0.7} 

\usepackage{caption}
\begin{document}

\title{The ridge integration method and its application to molecular sieving, demonstrated for gas purification via graphdiyne membranes}
\thanks{This is the pre-accepted version of the article published in \textit{Molecular Systems Design \& Engineering}, 2022, 7, 1622-1638. The final version is available at \href{https://doi.org/10.1039/D2ME00120A}{DOI: 10.1039/D2ME00120A}. © 2022 The Royal Society of Chemistry. /
Supplementary Information (ESI) available: [preliminary python code for ridge integration]. See the following GitHub repository: \url{https://github.com/hauser-group/ridge_integrator}}

\author{Christian W. Binder}
\affiliation{ 
Institute of Experimental Physics, Graz University of Technology, Petersgasse 16, 8010 Graz
}

\author{Johannes K. Krondorfer}
\affiliation{ 
Institute of Experimental Physics, Graz University of Technology, Petersgasse 16, 8010 Graz
}

\author{Andreas W. Hauser}%
\email{andreas.w.hauser@gmail.com}
\affiliation{ 
Institute of Experimental Physics, Graz University of Technology, Petersgasse 16, 8010 Graz
}

\date{\today}

\begin{abstract}
Eyring theory provides a convenient approximation to the rate of a chemical reaction as it uses only local information evaluated near extremal points of a given potential energy surface. However, in cases of pronounced anharmonicity and particularly low-lying vibrational frequencies, deviations from the correct reaction rate can become substantial. Molecular Dynamics simulations, on the other hand, are very costly at higher levels of theory, and of limited use since molecular reactions are `rare' events and hence statistically less accessible.

In this article, we present an alternative description for problems of gas separation and storage via two-dimensional materials such as porous graphene or flat metal-organic frameworks. Taking geometric advantage of the typical problem setting, our method is based on a statistical analysis of molecular trajectories near the so-called `ridge', a hypersurface which divides the reaction volume into a reactant and a product side. It allows for more realistic predictions of permeabilities and selectivities, e.g. derived from density functional theory, but without the considerable costs of a full molecular dynamics simulation on the corresponding Born-Oppenheimer potential energy surface. We test our method on the example of methane separation from nitrogen and carbon dioxide via a graphdiyne membrane.
\end{abstract}

\maketitle

\section{Introduction}
The ability to synthesize single-atom-thick materials\cite{Wang2021rev} such as porous graphene, graph\-diyne,\cite{Gao2019} and two-dimensional variants of metal organic frameworks\cite{Chakraborty2021} has triggered an avalanche of computational modeling attempts in recent years. Typically, molecular dynamics (MD) simulations are employed in order to describe the propagation of atoms, ions or molecules, either in solution or gas phase, through nanoporous structures. Although conceptually easy to grasp, and often simply based on a size-selectivity principle,\cite{book2016} determining realistic rates of pore propagation processes through two-dimensional materials is a challenge for several reasons.

First, the system of pore (stationary phase) and molecule (mobile phase) is typically rather large, and although the synthesis of periodically structured two-dimensional materials has become a recent reality, the actual propagation of the mobile phase remains a disordered, complex process with seemingly random trajectories. \cite{Ding2018} Second, the actual propagation of a single molecule through a nanopore, which can also be interpreted as a chemical reaction, is a `rare event' in terms of occurence. As a consequence, molecular dynamics calculations employing cost-effective force fields have become the well-established standard approach for theoretical estimates of permeabilities and gas selectivities. Third, when viewed as a chemical reaction, the reaction barrier, which corresponds to the event of a pore propagation, is extremely sensitive to the pore size,\cite{hauser2012b,hauser2012methane,book2016} the functionalization of the pore rim\cite{hauser2014c} and the correct treatment of long-range interactions.

Due to these issues it is a very time- and hence CPU-power-consuming endeavor to derive estimates for gas permeance and selectivity from statistics. For instance, a typical MD-simulation featuring a rather small system (a pore model and a single molecule) requires $\mathcal{O}(10^6)$ single point calculations. While importance sampling methods for rare events \cite{bolhuis2015practical} can offer a significant relief in this respect, even the most efficient variants demand the sampling of hundreds of trajectories to derive meaningful statistics. Assuming about $\mathcal{O}(10^2)$ to $\mathcal{O}(10^3)$ PES evaluations per trajectory, this amounts to a total expense between $\mathcal{O}(10^4)$ and $\mathcal{O}(10^5)$ single point evaluations, at best. Depending on the system of interest and the quality of the method, computational costs for one point evaluation can easily exceed one CPU hour, making high level methods such as density functional theory unfeasible for both of the above cases. Hence, lower level methods must be employed, leading to potentially significant errors in energy prediction.

An alternative is transition state (TS) theory, where a single trajectory is identified as a one-dimensional pathway that leads from the physisorption minimum on one side across the lowest transition state barrier to the physisorption minimum on the other side. The main cost required by this approach stems from the transition state search algorithms such as the nudged elastic band method,\cite{Henkelman2000,Henkelman2000a,Jonsson1998} the growing or frozen string method,\cite{Goodrow2009,Behn2011,Zimmerman2013} or other interpolation techniques which typically necessitate about $\mathcal{O}(10^2) - \mathcal{O}(10^3)$ single point evaluations. Further improvements can be achieved via machine learning methods.\cite{Peterson2016,Koistinen2017,Meyer2019} However, while allowing for treatments at a higher level of theory, the reduced picture of a single trajectory is, especially in the case of pore propagation problems, a rather questionable simplification. 

The objective of this article is to provide the reader with a new protocol, aiming for the best of both worlds: achieving the low computational cost of an Eyring-based calculation and therefore enabling the use of high level methods, while keeping the methodological accuracy of MD simulations. Our protocol is based on an extension of Eyring theory,\cite{eyring1935activated} but goes beyond the harmonic approximation through a cost-effective evaluation of the relevant partition sum factors. In our treatment, we fall back on the concept of the `ridge' as it has been introduced originally by Ionova and Carter in the context of an alternative TS search algorithm.\cite{Ionova1993,Miron2001} As described in these earlier works, the ridge has a dividing character for reactant and product volume. We note that elements of this idea can also be found in advanced methods such as transition interface sampling.\cite{bolhuis2015practical} We show that, for the typical case of molecular sieving, it is possible to identify a single interface or ridge. Within the simplification of a single particle as the mobile phase and a fully static, flat and a single-atomic-thick membrane, an excellent candidate for this dividing hyperplane is the actual geometrical plane of the latter. Taking advantage of the divisive property of the pore plane, we provide a protocol for the efficient computation of the relevant partition sums in a reduced subspace of the motional degrees of freedom. We provide a selection of numerical integration methods for the latter and find that a remarkably low number of single point evaluations is needed, surpassing trajectory- and even Eyring-based approaches in terms of computational efficiency. 
Furthermore, our protocol allows us to compare the predicitive performance of different levels of energy predictors in terms of chemical accuracy, revealing the methodological error of well known force fields in molecular sieving scenarios.

Picking the separation of CH$_4$ from N$_2$ and CO$_2$ via graphdiyne, a topical problem of natural gas purification\cite{Chawla2020} as a test case, we compare our ansatz with classical Eyring theory as well as full molecular dynamics simulations on the same potential energy surface (PES). The kinetic diameter of the two molecules differs by only 0.16~\AA{}, which makes this separation problem particularly challenging and very sensitive to details of the simulation such as the hydrophilic character and the polarity of a given porous structure.\cite{Chang2019,Nandanwar2020} Nitrogen gas is a significant impurity in natural gas which decreases its energy content. Currently, the standard technology for nitrogen removal from natural gas is cryogenic distillation, a costly and energy intensive process. At smaller scale, pressure swing adsorption has already become the mainstream technology for the purification of coal bed gas, where large losses of CH$_4$ are still an unsolved environmental problem and highly detrimental to the global climate due to the atmospheric greenhouse effect of methane.\cite{Hao2018} Membrane-based separation technologies, in particular molecular sieves, might allow for a less energy intense and therefore economically feasible separation.\cite{Carreon2017} Since permeance is inversely related to the membrane thickness, an effectively single-atom-thick membrane could be an excellent candidate for the task at hand.

\section{Methods}
\label{sec:Methods}
We start with a brief revision of Eyring theory to help pointing out the differences in comparison to our approach which is detailed further below. Semiclassical in nature, Eyring theory provides a convenient, approximative framework to determine chemical reaction rates.\cite{eyring1935activated,Evans1935} While the motion in the direction of the reaction coordinate is treated classically, all other directions are modeled by quantum theory. In its most general form, Eyring's equation for the prediction of reaction rates reads \cite{eyring1935activated}
\begin{equation}
    k=\frac{k_B T}{h}\frac{Z_{TS}}{\sum^{\text{min}}_i\left( Z_i \right)} \textrm{exp} \left( \frac{-\Delta E^\ddagger}{k_B T} \right),
    \label{eqn:1.11}
\end{equation}

where $Z_{i}$ and $Z_{TS}$ are partition sums evaluated at the corresponding minima and at the transition state of a given PES, respectively, and index $i$ runs over all vibrational degrees of freedom. In his original work, Eyring introduced harmonic approximations to evaluate this expression,\cite{eyring1935activated} but attempts have been made to generalize this strategy, e.g. by introducing anharmonic corrections.\cite{Anharmonic} 
For a better understanding of its limitations it is beneficial to derive the Eyring equation in a classical framework. We consider the case of a point-like particle in $n=3N$ dimensions, since any molecular structure consisting of $N$ atoms can be represented by such a system via a transformation to mass-weighted coordinates. The probability of finding this fictitious, classical particle, which represents a specific state of an arbitrary molecular many-body problem, within a given region $\Omega$ of phase space, is given by
\begin{equation}
P=\frac{1}{Z} \int_{\Omega} \exp \left(-\beta V\left(q_{1}, q_{2}, \cdots, q_{n-1}, q_{n}\right)\right) \prod_{i=1}^{n} \exp \left(-\beta{}\frac{p_{i}^{2}}{2 m}\right) d q_{i} d p_{i},
\label{prob_Phase}
\end{equation}
where $V(q_{1}, q_{2}, \cdots, q_{n-1}, q_{n})$ is denoting the potential energy as a function of $n$ mass-weighted coordinates $q_i$. Their corresponding momenta are $p_i$, $Z$ is the partition sum, and $\beta{}$ is an abbreviation for $(k_{\mathrm B}T)^{-1}$. We speak of a reaction event if the system rearranges its structure, in a continuous way, from a given local minimum geometry to another. In the course of such a transition, at some point a hyperplane in coordinate space is penetrated which separates the two neighboring minima. Usually, a plane located at the TS perpendicular to the reaction coordinate is introduced, with the latter being a one-dimensional parametrization of the minimum energy path that connects both minima via a TS. It is a fundamental postulate of Eyring theory that, once this separating hypersurface is crossed, the reactants will continue to form the product.\cite{jensen2017introduction} The reaction rate can now be defined as the probability $P_{\text{cross}}$ that the separating hyperplane will be crossed within a unit time interval
\begin{equation}
    k = \frac{P_{\text{cross}}(\delta t)}{\delta t}.
\end{equation}
$P_{\text{cross}}(\delta t)$ is also equal to the probability that a particle is located in a certain phase space volume. The latter shall be defined as the set of all phase space points which, if occupied, result in a transition event (an event of crossing the separating hyperplane) within a time interval $\delta t$. We will refer to this part of phase space as 'crossing volume'.  As is shown in the Appendix, by expanding the PES to second order at the TS, we arrive at
\begin{equation}
    P_{\text{cross}}(\delta t) \approx e^{-\beta \Delta E} \frac{\delta t}{Z} m^{\frac{1}{2}} \frac{2\pi}{ \beta^{2} \sqrt{k_{TS}}}
    \label{eqn:2.10_what}
\end{equation}
in the two-dimensional case, with $Z$ denoting the partition sum and $k_{TS}$ the eigenvalue of the Hessian at the TS. A completely analogous derivation in $n$ dimensions yields
\begin{equation}
    k \approx \frac{e^{-\beta \Delta E}}{Z} \frac{(2 \pi)^{n-1} m^{\frac{n-1}{2}}}{ \beta^n } \frac{1}{\prod_i^{n-1}{\sqrt{k_i^{\text{saddle}}}}},
    \label{EYRING_Z}
\end{equation}
where the $n$-th coordinate is identified as the reaction coordinate. The partition sum $Z$ can be approximated by a quadratic expansion at the minima. For the case of one minimum providing the dominant contribution, the expression condenses to
\begin{equation}
    k \approx e^{-\beta \Delta E} \frac{1}{2 \pi \sqrt{m}}  \frac{\prod_i^{n} \sqrt{k_i^{\text{min}}}}{\prod_i^{n-1}{\sqrt{k_i^{\text{saddle}}}}},
    \label{EYRING}
\end{equation}
with $k_i$ denoting the $i$-th eigenvalue of the Hessian evaluated at the minima and the TS structure respectively, excluding the single negative eigenvalue. The intrinsically local character of Eyring theory, however, leads to significant deviations from the correct reaction rate in cases of pronounced anharmonicity and particularly low-lying vibrational frequencies.\cite{ptavcek2018introduction}

\begin{figure}[H]
    \centering
    \includegraphics[width = 0.8\textwidth]{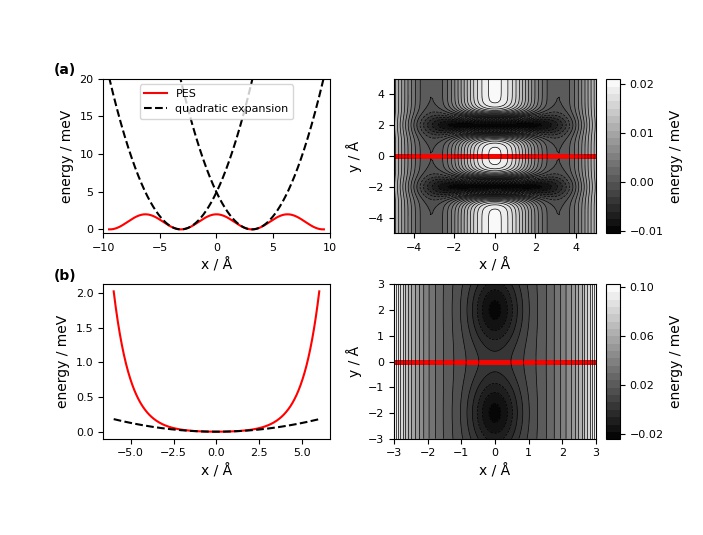}
    \vspace{-1cm}
    \caption{ Two cases illustrating the breakdown of Eyring theory via cuts (red line) through a two-dimensional PES at the transition state: (a) oscillatory, (b) fourth order parabola. The cut runs perpendicular to the reaction coordinate, which is either seen in projection (left) or from the top (right). Black dashed lines indicate harmonic approximations at the respective minima.}
    \label{fig:MODEL_RIDGE}
\end{figure}

The problem is illustrated in Figure~\ref{fig:MODEL_RIDGE}, where we cut through a fictitious two-dimensional PES at the TS state (red line in contour plots), exactly along the hyperplane perpendicular to the reaction coordinate. In case a), shown in the upper panel, the harmonic approximation is far too restrictive, as it does not account for possible, energetically almost equally feasible trajectories slightly off the optimum path along the reaction coordinate. This flaw can also be considered e.g. as a hindered rotation which is falsely interpreted as a vibration. A general, approximative \emph{ad hoc} correction for this type of failure has been suggested by Grimme,\cite{Grimme2012,Pracht2021} employing a continuous interpolation between partition sum contributions of rotational and vibrational degrees of freedom based on the frequency of the corresponding vibration.
Ribeiro \emph{et al.} have presented a similar ansatz for free energy calculations in solution.\cite{Ribeiro2011} Another related approach for the special case of surface adsorbates has been presented in Ref.~\citenum{Sprowl2016}. 

The opposite effect occurs in case b), illustrated in the lower panel of Figure~\ref{fig:MODEL_RIDGE}. Here, the harmonic approximation (dashed line) leads to an extremely flat parabola, while the actual PES is more closely resembling a box potential. This situation is typical for the usual pore models, where the box width is directly related to the pore diameter.

\subsection{Definition of the ridge}
In the above formulation we made use of a quadratic expansion of the PES at the TS to evaluate the probability that a point-like particle penetrates a plane perpendicular to the reaction coordinate within an infinitesimal time interval. A more general expression for the transition probability through a hyperplane $R$ with arbitrary orientation with respect to the coordinate system can be found by a similar strategy. This hyperplane is referred to as the `ridge' and is defined by the linear equation
\begin{equation}
    \tilde{R}(\mathbf{x}) = \mathbf{\hat{n}_R} \cdot \mathbf{x} - b = 0,
\end{equation}
with $\mathbf{\hat{n}_R}$ being the normal vector of the hyperplane $R$ and $b$ its distance to the origin. Again, a certain volume of phase space can be identified as the 'crossing volume' $\Omega(\delta{}t)$, which is given by the set of all points with a positive velocity $v_\perp = \mathbf{\hat{n}_R}\cdot \mathbf{p}$ perpendicular to the `ridge' which are within a distance of less than $-v_\perp \delta{}t$ in direction of the reaction coordinate,
\begin{equation}
    \Omega(\delta{}t) = \left\{ \mathbf{x},\mathbf{p} \in \mathbb{R}^n \; \vert \; \mathbf{\hat{n}_R} \cdot \mathbf{p} > 0,\;  \tilde{R}(\mathbf{x}) \in (-v_\perp\delta{}t,0)  \right\}. 
\end{equation}
All points contained in this volume cross the `ridge' with absolute certainty within a time interval $\delta{}t$. Without loss of generality we can transform into a coordinate system in which $x_n$ is the reaction coordinate and the `ridge' contains the origin, since the partition sum is invariant under rotation and translation of coordinates. In this case the `ridge' is defined by $x_n = 0$ and the 'crossing volume' is $\Omega(\delta{}t) = \left\{ \mathbf{x},\mathbf{p} \in \mathbb{R}^n \; \vert \; p_n > 0,\; x_n \in (-p_n \delta{}t,0)  \right\}$.  For an infinitesimal time step we can rewrite the integral in analogy to the derivation in the appendix and obtain 
\begin{equation} 
\begin{split}
P_{\text{trans}}(\delta t) = \frac{\delta t}{Z} \int_{-\infty}^{\infty} \int_{-\infty}^{\infty} ... \int_{-\infty}^{\infty} \prod_{i=0}^{n-1} d x_i e^{-\beta V(x_1, x_2,\dots,0)}  \\ \times \int_{-\infty}^{\infty} \int_{-\infty}^{\infty} ... \int_{0}^{\infty} p_{n} \prod_{i=0}^{n}  e^{-\beta p_{i}^2}  dp_i
\label{eqn:4.6}.
\end{split}
\end{equation}
Note the reduction of the lower integration limit from $-\infty$ to 0 in the reaction coordinate since any positive velocity will eventually lead to a transition.
A mass of one is assumed in Equation~\ref{eqn:4.6} since the problem will be formulated in mass weighted coordinates (see next section).

Before discussing the hyperplane $R$ or the `ridge' in the very obvious and easy interpretable case of a porous membrane, it should be mentioned that Equation \ref{eqn:4.6} can be derived analogously for a curved surface $R$ as long as the coordinate system is rotated such that, at any given point on $R$, the direction of $\mathbf{\hat{p}_n}$ is parallel to $\mathbf{\hat{n}_R}$. In this case, an additional Jacobi determinant appears in the integral due to the curvilinear coordinates. To be more concrete, we want to call a general hypersurface $R$ a `ridge' if it splits the PES between two minima in a sensible manner. It will be chosen in such a way that there is a probability close to one that a particle continues in the direction of the opposite minimum as soon as it crosses that `ridge'. Note that this definition is approximative and therefore less strict than in the formulation of Ionova and Carter.\cite{Ionova1993}   

\subsection{Relevance for molecular sieving}
\begin{figure}[H]
	\centering
		\includegraphics[scale=.45]{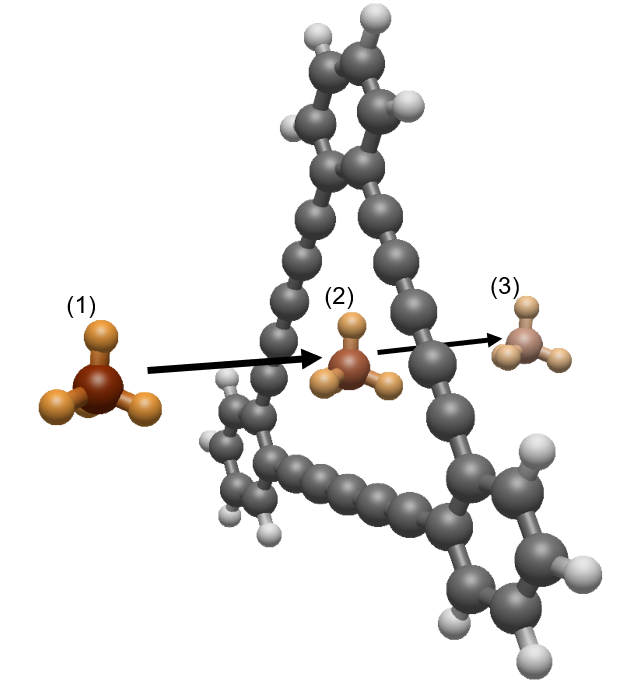}
	  \caption{Model system for the propagation of methane through a single pore of graphdiyne. The opposite adsorption minima are denoted as (1) and (3), the TS structure as (2). Note the terminating H atoms at the benzene rings replacing the neighboring chains of C atoms in the periodic 2D membrane structure.}\label{fig:TRANS_PORE}
\end{figure}

The propagation of a single molecule through a nanoporous membrane, from one side to the other, can be interpreted as a chemical reaction. When looking at Figure~\ref{fig:TRANS_PORE}, which depicts the molecular system to be studied in the next section, it becomes obvious that the concept of a ridge is particularly useful and easily interpretable for this type of problem: it is practically identical with the actual plane of the membrane, since it is reasonable to expect that the molecule will end up in the opposite minimum as soon as its center of mass moves across the pore plane.
The two flaws of the harmonic approximation intrinsic to Eyring theory (see Figure~\ref{fig:MODEL_RIDGE}) are obvious for this very specific problem of molecular sieving. First, due to symmetry, the combined system of pore and molecule exhibits multiple equivalent saddle points, leading to a rather complicated energy landscape with respect to the rotational degrees of the molecule of interest. Second, while comparably weak van der Waals forces are dominating the interaction between the pore and the molecule near the TS state, the membrane is fully impermeable otherwise, which renders the PES rather box-like in the two translational degrees of freedom of the molecule which are perpendicular to the reaction coordinate.  

\subsection{Choice of coordinates} 
While coordinates in the usual abstract 3$N$-dimensional Cartesian space,

\begin{equation}
\textbf{x} = (x_1,y_1,z_1,x_2,y_2,z_2, ...,x_n,y_n,z_n),
\end{equation}

\noindent and their mass-weighted equivalents,

\begin{equation}
\textbf{q} = (x_1 \sqrt{m_1},y_1 \sqrt{m_1},z_1 \sqrt{m_1}, x_2 \sqrt{m_2},..., x_n \sqrt{m_n}, y_n \sqrt{m_n}, z_n \sqrt{m_n}),
\end{equation}

\noindent have been used in parts of the derivation for the classical Eyring theory, final expressions for the rate equation in the case of pore propagation problems are formulated in a set referred to as TRV-coordinates, with TRV denoting translational, rotational and vibrational degrees of freedom. This nomenclature accounts for the fact that rotational and translational degrees of freedom of the mobile phase, i.e. the molecule attempting a pore propagation, are crucial for the description of the process, while internal vibrational degrees of freedom and vibrational motion of the pore are of less importance. Consequently, the TRV coordinates incorporate only degrees of freedom of the mobile phase, i.e. the pore is treated as a static object from now on, appearing as an external potential to the propagating molecule. For the sake of readability, we keep the symbol $n = 3N$, but note that $N$ is now denoting the number of atoms of the propagating molecule only. Besides pore movement, also internal vibrations of the mobile phase are going to be neglected in all partition sums. In other words, we are implicitly assuming that these contributions are not varying during propagation.

A vector in the TRV-coordinates will be denoted by the symbol $\boldsymbol{\xi}$ and its coordinates by $\xi_i$. Since we will often refer to a specific $\xi_i$ we define $\boldsymbol{\xi} = (t_1,t_2,t_3,r_1,r_2,r_3,v_1,v_2,...,v_{n-6})$. Here, $t_i$ and $r_i$ label the translational and rotational degrees of freedom, while $v_i$ labels the vibrational ones. The TRV-coordinates are of curvilinear nature, hence 
\begin{equation}
    \xi_1 = \xi_1(q^1, q^2,..., q^n),\, \xi_2 = \xi_2(q^1, q^2,..., q^n),...,  \xi_n = \xi_n(q^1, q^2,..., q^n). 
\end{equation}
The local basis vectors $\boldsymbol{l}^{TRV}_i$ in these coordinates can be computed in the usual way
\begin{equation}
\boldsymbol{\tau_l}=\left(\begin{array}{c}
\frac{\partial q_{1}}{\partial t_{l}} \\
\frac{\partial q_{2}}{\partial t_{l}} \\
\vdots \\
\frac{\partial q_{n}}{\partial t_{l}}
\end{array}\right) \quad \boldsymbol{\rho_{j}}=\left(\begin{array}{c}
\frac{\partial q_{1}}{\partial r_{j}} \\
\frac{\partial q_{2}}{\partial r_{j}} \\
\vdots \\
\frac{\partial q_{n}}{\partial r_{j}}
\end{array}\right) \ldots \quad \boldsymbol{\nu_{k}}=\left(\begin{array}{c}
\frac{\partial q_{1}}{\partial v_{k}} \\
\frac{\partial q_{2}}{\partial v_{k}} \\
 \\
\frac{\partial q_{n}}{\partial v_{k}}
\end{array}\right).
\end{equation}

For a molecule that is not subject to any external potential, the normalized TRV basis vectors $\hat{\boldsymbol{\tau_l}}$, $\hat{\boldsymbol{\rho_j}}$ and $\hat{\boldsymbol{\nu_k}}$ correspond to the eigenvectors of the Hessian at a given point. Since the translational and rotational eigenvectors share zero as their respective eigenvalue, there is a certain freedom in their choice.\cite{ochterski1999vibrational}
For the translational coordinates we choose center-of-mass coordinates ($t_1 = x_{com}, \quad t_2 = y_{com}, \quad t_3 = z_{com}$). The rotational coordinates are chosen as the Euler angles with the axes of rotation also passing through the molecular center of mass.

\subsection{Ridge integration method for molecular sieving}
Orienting the membrane along the xy-plane, the ridge can be approximated by the equation $t_3 = \xi_3 = 0$. Furthermore, we can clearly identify $t_3$ as the reaction coordinate at the transition state. This allows to carry out the integrals of Equation \ref{eqn:4.6} in TRV-coordinates, yielding

\begin{equation} 
\begin{aligned}
P_{\text{trans}}(\delta t) &= \frac{\delta t}{Z}\int_{-\infty}^{\infty} \int_{-\infty}^{\infty} ... \int_{\-\infty}^{\infty} \prod_{i=1 \atop i \neq 3}^{n} d \xi_i e^{-\beta V(t_1, t_2,t_3 = 0,r_1,r_2,r_3,v_1,...,v_{n-6})} \\
&\times \text{det}(S_{TRV}(t_1,t_2,r_1,r_2,r_3,v_1,...,v_{n-6})) \\
&\times \int_{-\infty}^{\infty} \int_{-\infty}^{\infty} \int_{0}^{\infty} \int_{-\infty}^{\infty} ... \int_{-\infty}^{\infty}  p_{\xi_3} \prod_{i=0}^{n}  e^{-\beta p_{\xi_i}^2}  dp_{\xi_i}
\label{eqn:4.7},
\end{aligned}
\end{equation}

\noindent where the determinant of 

\begin{equation}
S_{TRV} = \left(\boldsymbol{\tau_1} \enspace \boldsymbol{\tau_2} \enspace \boldsymbol{\hat{\tau_3}} \enspace \boldsymbol{\rho_1} \enspace \boldsymbol{\rho_2} \enspace \boldsymbol{\rho_3} \enspace \boldsymbol{\nu_1} \ldots \boldsymbol{\nu_{n-6}} \right)
\label{eqn:STRV}
\end{equation}

\noindent is again the surface area of an infinitesimal ridge element, with $\boldsymbol{\hat{\tau_3}}$ denoting the unit vector in $\tau_3$ direction. The partition sum $Z$ is given by

\begin{align} 
\begin{split}
Z &= \int_{-\infty}^{\infty} \int_{-\infty}^{\infty} ... \int_{\-\infty}^{\infty} \prod_{i=1}^{n} d \xi_i e^{-\beta V(t_1, t_2,t_3,...,v_{n-6})} \\ &\quad\times \text{det}(J_{TRV}(t_1,t_2,t_3,r_1,...,v_{n-6})) \int_{-\infty}^{\infty} \int_{-\infty}^{\infty} ... \int_{-\infty}^{\infty} \prod_{i=0}^{n}  e^{-\beta p_{\xi_i}^2}  dp_{\xi_i},
\label{partsum}
\end{split}
\end{align}

\noindent where $J_{TRV}$ is the Jacobian in $TRV$-coordinates,

\begin{equation}
J_{TRV} = \left(\boldsymbol{\tau_1} \enspace \boldsymbol{\tau_2} \enspace \boldsymbol{\tau_3} \enspace \boldsymbol{\rho_1} \enspace \boldsymbol{\rho_2} \enspace \boldsymbol{\rho_3} \enspace \boldsymbol{\nu_1} \ldots \boldsymbol{\nu_{n-6}} \right).
\label{eqn:JTRV}
\end{equation}

We employ the notation used in Equation \ref{eqn:JTRV} to refer to a matrix with its columns being identical to the featured vectors $\boldsymbol{l}^{TRV}_i$. In Equations \ref{eqn:4.7} to \ref{partsum}, $V(t_1, t_2,t_3,...,v_{n-6})$ is a superposition of the external potential felt by the molecule due to the pore and the intramolecular interaction potential. In a first approximation, we use the potential generated by the relaxed pore geometry and neglect any pore movement. Futhermore, we demonstrate in the Appendix that, after partial evaluation of the integrals, Equation \ref{eqn:4.7} reduces to

\begin{equation}
\begin{aligned}
{P}_{\text{trans}}(\delta t) \approx & \frac{\delta t}{Z_{TR}} \int_{\Omega} d t_{1} d t_{2}  d r_{1} d r_{2} d r_{3} e^{-\beta V\left(t_{1}, t_{2}, t_{3}=0, r_{1},r_{2}, r_{3}\right)} \\
& \times \text{det}(S_{\text{TRV}}(t_1,t_2,r_1,r_2,r_3)) \frac{1}{\sqrt{2 \pi \beta}}.
\label{RIDGERR}
\end{aligned}
\end{equation}

This step assumes the validity of the rigid rotor approximation (see Appendix \ref{A2}), which requires the molecule to only weakly change its shape during the process of interest. If this is the case, the $\nu$-dependence of $S_{TRV}$ can be neglected and, as denoted in Equation \ref{RIDGERR}, it then effectively depends on the rotational and translation coordinates only. Consequently, in this combined `frozen' pore and molecule approximation, the relevant phase space shrinks considerably as it comprises only the six remaining dimensions of molecular translation and rotation. The reduced partition sum $Z_{TR}$ is now given by

\begin{equation}
\begin{aligned}
Z_{TR}= &\int_{\Omega} d t_{1} d t_{2} d t_{3} d r_{1} d r_{2} d r_{3} e^{-\beta V\left(t_{1}, t_{2}, t_{3}, r_{1},r_{2}, r_{3}\right)} \\
&\times \text{det} (J_{\text{TRV}}(t_1,t_2,t_3,r_1,r_2,r_3)).
\label{ZRT}
\end{aligned}
\end{equation}
The integral in Equation \ref{RIDGERR} will be referred to as ridge integral in the following sections.

In the case of linear molecules, we define two rotational coordinates $r_1$ and $r_2$ instead of three when calculating the transition probability. The final expression for ${P}_{\text{trans}}$ then reads

\begin{equation}
\begin{aligned}
{P}_{\text{trans}}(\delta t) \approx & \frac{\delta t}{Z_{TR}} \int_{\Omega} d t_{1} d t_{2}  d r_{1} d r_{2} e^{-\beta V\left(t_{1}, t_{2}, t_{3}=0, r_{1}, r_{2}\right)} \\
& \times \text{det}(S_{\text{TRV}}(t_1,t_2,r_1,r_2)) \frac{1}{\sqrt{2 \pi \beta}},
\label{P_trans}
\end{aligned}
\end{equation}
where
\begin{equation}
Z_{TR}= \int_{\Omega} d t_{1} d t_{2} d t_{3} d r_{1} d r_{2} e^{-\beta V\left(t_{1}, t_{2}, t_{3}, r_{1}, r_{2}\right)} \text{det} (J_{TRV}(t_1,t_2,t_3,r_1,r_2)).
\label{redZ_TR}
\end{equation}

\subsection{Efficient evaluation of partition sums}

The last remaining step in our quest to determine the transition probability is now to evaluate Equations \ref{RIDGERR} and \ref{ZRT} numerically. Via a relocation of the problem into the roto-translational subspace we achieve two things. The integrand variance as well as the dimensionality of the problem are quenched, achieved by the elimination of oscillatory degrees of freedom. This drastically reduces the number of single point evaluations required for the numerical integration of Equations \ref{RIDGERR} and \ref{ZRT}. The latter can be done using quadrature rules of the form

\begin{equation}
    I(f) := \int_{\Omega} f(\boldsymbol{x}) \;\mathrm{d} x \approx \sum_{i=1}^N w_i f(\boldsymbol{x}_i) =: Q(f),
    \label{quadrature}
\end{equation}
with weights $\{w_i\}$ and abscissas $\{\boldsymbol{x}_i\}$. Here $f$ represents the integrand of interest (in our case either that of Equation \ref{RIDGERR} or \ref{ZRT}), which is of the form

\begin{equation}
f(\boldsymbol{x})=  e^{-\beta V(\boldsymbol{x})} \text{det} J(\boldsymbol{x}),
\label{f}
\end{equation}
where $\boldsymbol{x}$ represents the integration coordinates of the given system and $J$ the respective Jacobian.
In principle the integration may be carried out over a grid (trapezoidal rule, midpoint rule, etc.). However, we do not expect this strategy to be feasible, since too many single point evaluations are needed to resolve the integrand this way. Hence, we will assess the performance of two more suitable numerical integration routines, Monte Carlo Integration and $l$1-quadrature, instead.

\subsubsection{Monte Carlo Integration}
\label{sec:mcis}

For integrands exhibiting low variance we identify Monte Carlo integration as a suitable method since the error of an integral calculated via the latter is directly linked to the integrands variance  \cite{weinzierl2000introduction}, which can be estimated via

\begin{equation}
\sigma^{2}(f)=\frac{1}{N-1} \sum_{i=1}^{N}\left(f\left(\boldsymbol{x}_{i}\right)- \langle f \rangle \right)^{2}.
\end{equation}
In this case, the weights $\{w_i\}$ in Equation \ref{quadrature} are chosen to be $\frac{1}{N}$, where $\{\boldsymbol{x}_{i}\}$ values are drawn from a uniform distribution over the entire integration domain. 
However, due to the exponential nature of the integrand, the variance is still significant. While a simple Monte Carlo integration yields acceptable results, the integrals at hand are an excellent target for further improvement through importance sampling. Rewriting Equation \ref{quadrature} as 

\begin{equation}
    I(f) = \int_{\Omega} \frac{f(\boldsymbol{x})}{P(\boldsymbol{x})} P(\boldsymbol{x}) \;\mathrm{d}x,
    \label{expandedInt}
\end{equation}
the integral $I$ can now be evaluated differently. For one, the abscissas $\{\boldsymbol{x}_{i}\}$ are sampled according to a probability distribution $P(\boldsymbol{x})$. For the other, the original function $f(\boldsymbol{x})$ is replaced by $\frac{f(\boldsymbol{x})}{P(\boldsymbol{x})}$ while the weights are again given by $\frac{1}{N}$.

Consequently, it is advantageous to choose $P(\boldsymbol{x})$ such that the variance of the new integrand is minimal. To achieve the latter, it is beneficial to choose a $P(\boldsymbol{x})$ that mimics $f(\boldsymbol{x})$. An excellent choice for $P(\boldsymbol{x})$ is a normalized lower level approximation to $f$. This can be realized by determining $V(\boldsymbol{x})$ with a lower cost theory (e.g. cost effective force fields). Equation \ref{expandedInt} then becomes

\begin{equation}
    I(f) = \int_{\Omega} e^{-\beta (V_{hl}(\boldsymbol{x}) - V_{ll}(\boldsymbol{x}))} e^{-\beta V_{ll}(\boldsymbol{x})} \det J(\boldsymbol{x})\;\mathrm{d}x,
    \label{ll_hl_approx}
\end{equation}
where $V_{ll}$ and $V_{hl}$ are the respective potentials determined by the low level and high level theory.
Obviously, the computational cost is significantly reduced if the high level potential is well approximated by its low level counterpart.

\subsubsection{Integration via $l$1-quadrature}
\label{sec:l1}

Another popular integration routine is Gauss quadrature, where abscissas and weights are chosen such that a certain set of basis functions, e.g. polynomials, are integrated exactly. The convergence of the numerical integration is thereby done by increasing the set of basis functions. A suitable choice for polynomial basis functions are the relevant Taylor polynomials up to a certain degree in multiple dimensions. Due to its exponential character, it is difficult to approximate the function $f(\boldsymbol{x})$ with polynomials. In other words, a large basis, and thus a large number of point evaluations, is required to resemble the function $f$ and yield a proper estimate for the value of the integral. 
The function $\frac{f(\boldsymbol{x})}{P(\boldsymbol{x})}$, however, shows a much smoother behavior as long as the change in the difference $(V_{hl}(\boldsymbol{x}) - V_{ll}(\boldsymbol{x}))$ remains small. As long as the approximating low level method is resembling the high level method sufficiently well, the function  $e^{-\beta (V_{hl}(\boldsymbol{x}) - V_{ll}(\boldsymbol{x}))}$ is well represented by a small polynomial basis. Thus, by interpreting the first exponent in Equation \ref{ll_hl_approx} as the integrand and the term $e^{-\beta V_{ll}(\boldsymbol{x})} \det J(\boldsymbol{x})$ as the (non-negative) integration measure, we can construct a numerical integration scheme that integrates the difference of the low level and the high level method.

The $l$1-quadrature technique has proven to be a very suitable for this purpose in our case. Here, the integral is also approximated by Equation \ref{quadrature}, but the abscissas are determined by a two-step-process. Similar to the case of Monte Carlo Importance sampling, the first step consists of a random sampling of points according to the integration measure, ensuring that the abscissas are located in relevant areas.\cite{l1Quad} The second step is a further refinement of the drawn abscissas while choosing the weights such that a given set of basis functions is integrated exactly using a minimum number of the proposed abscissas.\cite{l1Quad2} This can be achieved by sieving the drawn abscissas and determining the corresponding weights via a simplex algorithm (details can be found in Appendix~\ref{A:l1_quadrature}), ensuring that the maximum number of abscissas is equal to the number of basis functions. We use a multidimensional polynomial basis up to a certain degree. A big advantage of this integration technique is that the abscissas can be chosen to be nested, meaning that the abscissas for a lower degree are contained in the abscissas for a higher degree. This proves extremely useful, since convergence of the numerical integration can be obtained without additional point evaluations.

\section{Results}
\label{sec:Ass_ridge}
\begin{figure}[H]
    \centering
    \includegraphics[width = 0.4\textwidth]{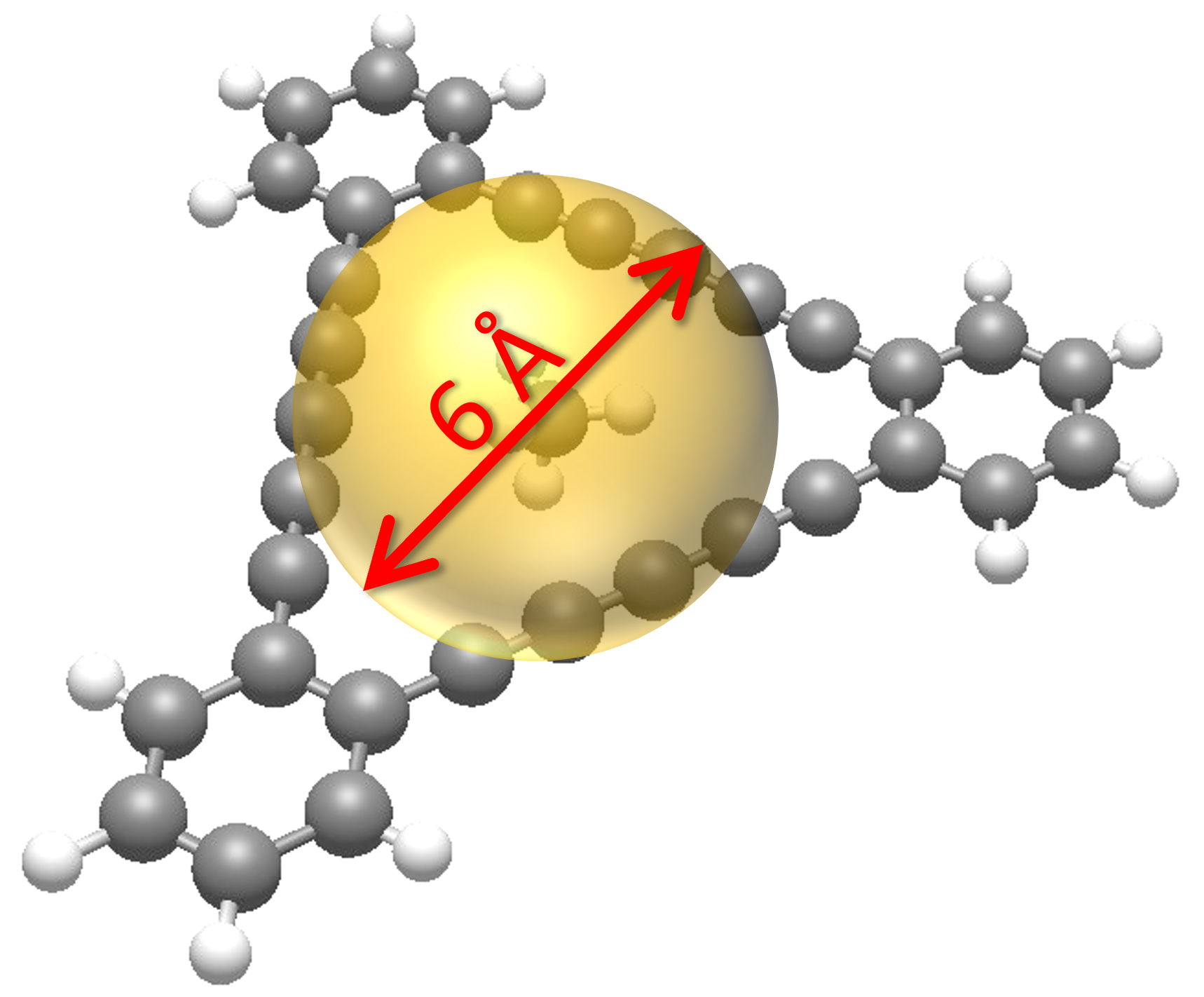}
    \textbf{}
    \caption{General setup of the molecular dynamics simulation. The molecule of interest is confined to the yellow sphere. As soon as a selected atom of the mobile phase attempts to leave the sphere, it is pushed back inside by a harmonic potential.}
    \label{fig:MOLDYN}
\end{figure}

The ridge integration method is now validated for the topical problem of natural gas or bio-gas purification via membrane technologies. We investigate the propagation of its main constituents, methane, carbon dioxide and nitrogen, through a single pore of a graphdiyne sheet (see Figure \ref{fig:MOLDYN}).

Our study involves several steps, each of them described in a separate subsection. After a brief overview of our computational pore model, it starts with
\begin{itemize}
    \item  a comparison of either predicted or counted numbers of pore transition events for the various gas species and methods applied. The former applies to TS-based methods, the latter to MD-based simulations. Note that this first step does not claim to produce realistic data, since all methods will rely on the same energy calculator, i.e. the same underlying potential energy surfaces for the sake of a fair comparison. This calculator will be the GFN-FF force field (FF) of the Grimme group,\cite{Grimme2019} which is of sufficiently low computational cost to allow for full MD simulations. Within the context of this first comparison, the results of the molecular dynamics simulation will be considered `exact' and serve as a reference when evaluating the predictive power of Eyring theory and the proposed ridge method. This step is addressed in Sections \ref{sec:model_pore} and \ref{sec:graphdiyne}.
    \item In a second step (Section~\ref{sec:cost_acc}), we will discuss the computational costs of all methods and compare their accuracy. Most relevant is the number of single point calculations needed in order to retrieve meaningful estimates.
    \item Having demonstrated the improvements in accuracy for the ridge method in comparison to Eyring Theory, another question to ask is whether low-level energy predictors yield sufficiently accurate results in order to obtain reasonable transition rates in problems of pore propagation. In other words, can a full MD simulation, feasible only for low-cost methods, be considered `good enough' for realistic predictions. The answer is no, as will be demonstrated in Section~\ref{sec:Methval}.
    \item Finally, in step~4, we provide realistic selectivity predictions obtained with DFT in combination with the ridge method and present our results in Section~\ref{sec:gassep}.
\end{itemize}

\subsection{Model pore and computational details} \label{sec:model_pore}
Pore propagation can be thought of as a three stage process, starting with the adsorption onto one side of the membrane, followed by a penetration of the pore plane, and ending with the escape of the  molecule back into the continuum on the opposite side of the membrane. Consequently, one can attribute three reaction rate constants k$_1$, k$_2$ and k$_3$ to these processes, respectively. Assuming that advent and escape are diffusion-driven processes occurring at a much higher rate than pore transition, k$_2$ is clearly rate-determining and therefore of main interest. In order to be able to simulate the corresponding process via molecular dynamics, a constraining potential is introduced, in the form of a spherical volume of radius $r=3$\,\AA{} in the case of N$_2$ and CH$_4$, and of radius $r=4$\,\AA{} in the case of CO$_2$. The size of the confining sphere, centered at the pore center, is chosen to fully accommodate the adsorption minima of the corresponding molecule. Consequently, twice the distance between pore center and adsorption minimum was used as the sphere radius. In all three cases this radius is also small enough to only allow for transition events through the pore center. A harmonic potential is applied to keep the mobile phase within the confining sphere. 

The number of expected transition events per ns is calculated with our method (Equation \ref{RIDGERR}) and via Eyring theory (Equation \ref{EYRING_Z}), followed by a comparison to the MD simulation benchmark. In order to determine this number, we store the $z$-position of the center of mass of the respective molecule every 200~fs. Each time the difference between that stored position and the $z$-component of the pore center changes sign, a transition event is counted. The discrete nature of this checking process automatically avoids the counting of faster recrossing events related to incomplete transitions. All three methods, ridge integration, Eyring and full molecular dynamics, employ the same FF-based single point calculator. The temperature is controlled via a Langevin thermostat as it is implemented in the atomic simulation environment (ASE),\cite{ASE} a convenient Python library used as an interface between the XTB program package\cite{grimme2017robust} and our own Python code. All examinations are conducted for temperatures ranging from 100 to 600~K. To investigate the impact of pore vibration, all calculations are done for a static pore at first and then repeated for a more flexible geometry. In the former, all pore atoms are locked to their equilibrium positions during simulation; in the latter, only the terminating H atoms of the pore are kept fixed.

\subsection{Graphdiyne pore transitions per time} \label{sec:graphdiyne}
Since the ridge method is currently implemented only for `frozen' pore scenarios, i.e. for geometrically stiff pores, we begin our investigation with this case. For the calculation of the Hessian (see Equation \ref{EYRING_Z}) we consider only the subspace of molecular degrees of freedom of the mobile phase. The simulations cover a timespan of 1.5 to 10~ns.

\subsubsection{Frozen pore}
\begin{figure}[H]
     \centering
     \begin{subfigure}[b]{0.32\textwidth}
         \centering
        \includegraphics[width = \textwidth]{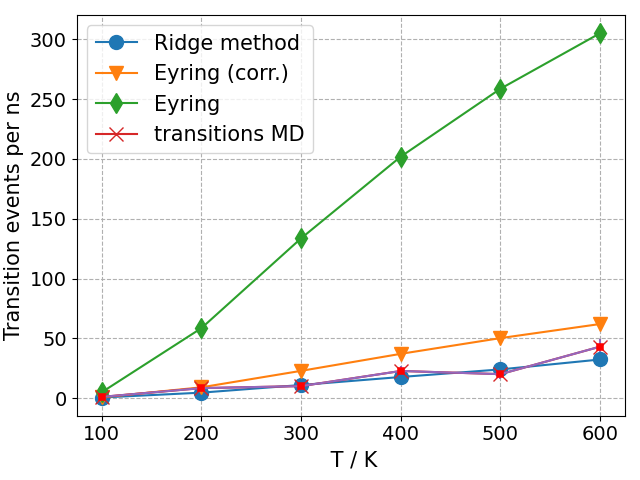}
        \caption{Methane}
        \label{fig:comp-CH4_frozen}
     \end{subfigure}
     \begin{subfigure}[b]{0.32\textwidth}
         \centering
         \includegraphics[width = \textwidth]{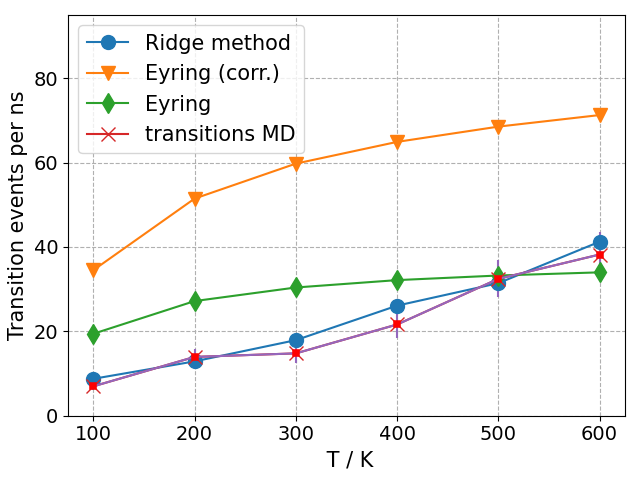}
        \caption{Molecular nitrogen}
        \label{comp-N2_frozen}
     \end{subfigure}
     \begin{subfigure}[b]{0.32\textwidth}
         \centering
         \includegraphics[width = \textwidth]{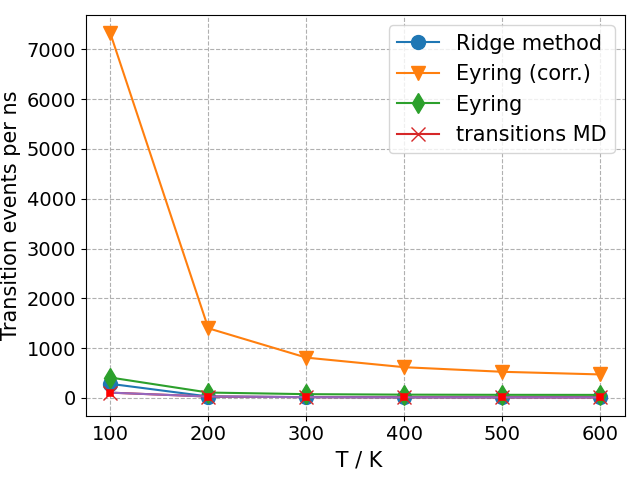}
        \caption{Carbon dioxide}
        \label{fig:comp-CO2_frozen}
     \end{subfigure}
        \caption{Predicted and observed transition events per ns, obtained for N$_2$, CH$_4$ and CO$_2$ propagating through a graphdiyne model pore.}
        \label{fig:reac-rate-comp}
\end{figure}

We start with the propagation of CH$_4$. A direct comparison of the MD results and the prediction methods is given in Figure~\ref{fig:comp-CH4_frozen}. Uncertainties for the simulations, depicted by errorbars for the MD data points, have been determined statistically. The remaining relative uncertainty of the observed transition events lies at around 10\,\%. Besides the predictions made by ridge integration, the graph also contains the estimates of two variants of Eyring Theory. The first curve, simply referred to as `Eyring' in the legend, is obtained through a straightforward implementation of Equation~\ref{EYRING}. The second, denoted as `Eyring (corr.)', applies a correction as it has been suggested in Ref.~\citenum{Grimme2012}, offering an effective compensation of problematic motional degrees of freedom such as hindered rotations. It is based on a continuous interpolation between rotational and vibrational contributions to the entropy for low frequency modes. We stick to the parameters as recommended in the original publication of Grimme;\cite{fn5} the only adjustment we make is to use classical instead of quantum mechanical partition sums for the calculation of the vibrational entropy contributions. This way we ensure comparability of the theoretical predictions and the classical molecular dynamics simulations. The figure illustrates a pronounced overestimation of transition events through Eyring theory, which is significantly improved by the entropy-interpolation ansatz, while the ridge integration method shows almost perfect agreement with the simulated counts.

The propagation of N$_2$ is analyzed in Figure~\ref{comp-N2_frozen}. Again, the ridge integration method is almost spot on with the MD simulation. Interestingly, while Eyring theory also provides a reasonable estimate, the entropy-corrected Eyring prediction overestimates the count of transition events in this case. Unlike for the highly symmetric CH$_4$, where a partial replacement of vibrational with rotational entropy contributions proved useful, this strategy appears to be less effective for linear molecules at the relevant extreme points. Furthermore, since adsorption minima and transition state are both affected by anharmonicities, such an \emph{ad hoc} correction can tilt the highly sensitive outcome either way, as it is affecting partition sums in the denominator ($Z$) as well as in the numerator of Equation~\ref{EYRING}. 

Finally, in Figure~\ref{fig:comp-CO2_frozen}, we compare the predictions for the propagation of CO$_2$ through the graphdiyne pore. This scenario yields an entirely different outcome, which originates from the fact that the propagation of CO$_2$ through graphdiyne is barrierless: the corresponding intrinsic reaction path, when calculated at the GFN-FF level, shows an adsorption minimum at the center of the pore, with two TS structures of negligible height separating the two adsorption minima from the local minimum inside the pore. While this renders the two Eyring-based approximations meaningless, the theoretical machinery can still be applied. However, note the substantial deviation (about two orders of magnitude) produced by the entropy-corrected Eyring ansatz in this ill defined situation. The ridge integration method, on the other hand, is unaffected by the non-existence of a rate-determining TS structure as long as there exists a valid definition of the ridge. Its prediction agrees with the simulation within 20\% for temperatures of 200~K and above. Only at temperatures as low as 100~K the ridge integration method shows a significant deviation from the simulation due to the multiple counting of incomplete transitions of the CO$_2$ molecule. This is an inevitable side effect of the `confining' nature of the adsorption minimum inside the pore at sufficiently low temperatures. In this regime, the number of counted transitions becomes strongly dependent on the position update interval chosen in the MD simulation (200~fs), and any Eyring-based theory will overshoot by definition due to its intrinsic assumption of a complete transition for every crossing.

\begin{table}[H]
\centering
\caption{Predicted and observed numbers of transition events per ns for N$_2$, CH$_4$ and CO$_2$.} 
\begin{tabular}{@{}lcccccccccc@{}}
\toprule
\multicolumn{1}{c}{CH$_4$ - Temperature (K)} & 100 & 200 & 300 & 400 & 500 & 600 &  \\ \midrule
\textbf{Eyring theory}  & 4.91 &  58.52 & 133.63 & 201.94 & 258.70 & 305.16   \\
\textbf{Eyring (corr.)}     &  0.63 &  9.00 & 22.77 & 37.00 & 50.15 & 61.94  \\
\textbf{Ridge integration}        & 0.47 & 4.54 &  10.93& 17.63 & 23.96& 32.27 &   \\
\textbf{MD benchmark}          & 1.00 &  8.33 & 10.00   &  22.67 & 20.00 &      43.00 &  \\ \hline

\multicolumn{1}{c}{N$_2$ - Temperature (K)} & 100 & 200 & 300 & 400 & 500 & 600 &  \\ \midrule
\textbf{Eyring theory}  & 19.43 & 27.18 & 30.40 & 32.15 & 33.24 & 33.99  \\
\textbf{Eyring (corr.)}     & 34.51 & 51.48 & 59.77 & 64.92 & 68.54 & 71.28   \\
\textbf{Ridge integration }           & 8.75 & 12.89 & 17.92 & 26.100 & 31.43 & 41.26 &   \\
\textbf{MD benchmark}               & 6.97 & 13.94 &  14.79 & 21.63 & 32.45 & 38.22 &   \\ \midrule

\multicolumn{1}{c}{CO$_2$ - Temperature (K)} & 100 & 200 & 300 & 400 & 500 & 600 &  \\ \midrule
\textbf{Eyring theory}              & 410.64 & 108.84 & 76.29 & 66.64 & 63.02 & 61.74   \\
\textbf{Eyring (corr.)}    & 7323.03 & 1399.03 &  809.66 &  617.39 & 525.44 &  472.32   \\
\textbf{Ridge integration}         & 283.72 & 29.38 & 17.16 & 13.97 & 14.01 & 15.59   \\
\textbf{MD benchmark}          & 145.06   & 34.97 & 20.37   &  25.51 & 15.43 &      14.81 &  \\ \hline\bottomrule
\end{tabular}
\end{table}

\subsubsection{Unconstrained pore}
\begin{figure}[H]
     \centering
     \begin{subfigure}[b]{0.33\textwidth}
         \centering
         \includegraphics[width = \textwidth]{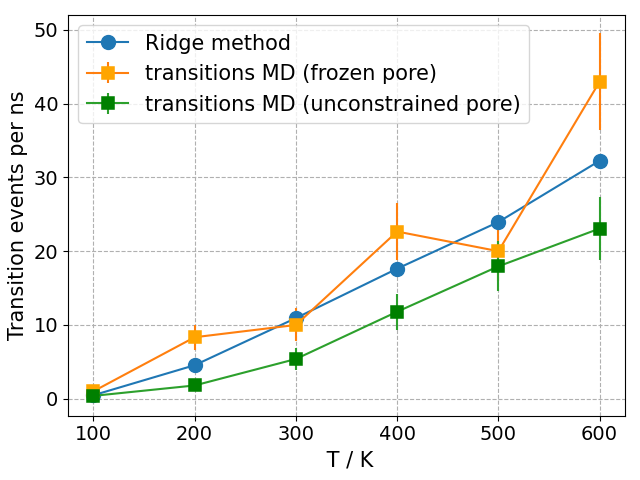}
        \caption{Methane}
        \label{fig:Move_vs_frozen}
     \end{subfigure}%
     \begin{subfigure}[b]{0.33\textwidth}
         \centering
         \includegraphics[width = \textwidth]{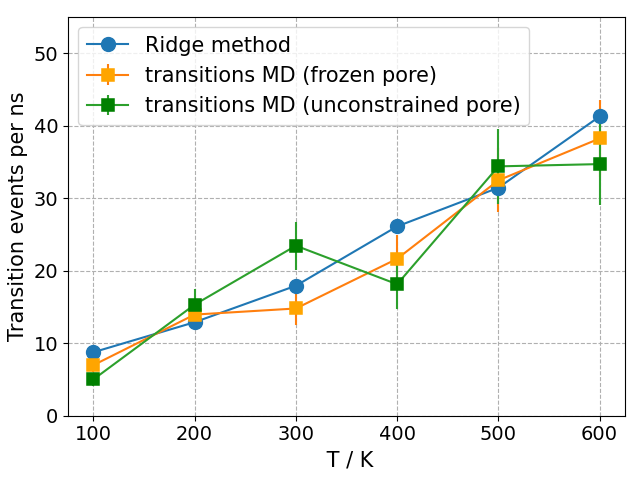}
        \caption{Molecular nitrogen}
        \label{fig:Move_vs_frozen_N2}
     \end{subfigure}%
     \begin{subfigure}[b]{0.33\textwidth}
         \centering
         \includegraphics[width = \textwidth]{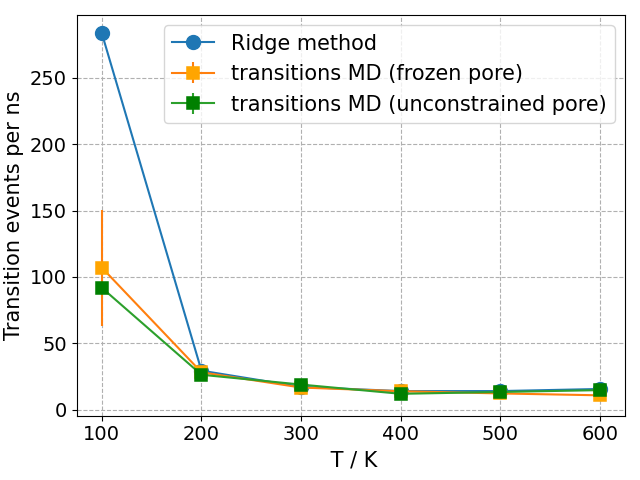}
        \caption{Carbon dioxide}
        \label{fig:Move_vs_frozen_CO2}
     \end{subfigure}
        \caption{Number of MD transition events per ns for moving and frozen pore cases plotted as a function of temperature, and compared to the ridge-based prediction.}
        \label{fig:kjdglajksgb}
\end{figure}

Since an unconstrained pore represents a more realistic case, we want to compare this computationally more challenging scenario to the results obtained for the frozen pore. The model system is identical to the structure shown in Figure~\ref{fig:MOLDYN}, but now only the bond-saturating H atoms at the benzene rings are kept at fixed positions during the MD-simulation. A comparison of the numerical results is provided in Table~\ref{tab:constr} for various temperatures. Figures~\ref{fig:Move_vs_frozen} to~\ref{fig:Move_vs_frozen_CO2} show transition event predictions of the ridge method and of the MD simulations performed for the fully frozen as well as the unconstrained pore. 

As can be seen, excellent agreement of the number of observed transitions per ns for frozen and moving pore cases can be confirmed for all three molecular species. From this we conclude that the ridge integration method, although limited to frozen pore models, should yield reasonable estimates for the reaction rates of flexible, and hence more realistic porous membranes.

\begin{table}[H]
\centering
\caption{Predicted and observed numbers of transition events per ns, for CH$_4$, N$_2$ and CO$_2$, comparing constrained and unconstrained pore models.} \label{tab:constr}
\begin{tabular}{@{}lcccccccccc@{}}
\toprule
\multicolumn{1}{c}{CH$_4$ - Temperature (K)} & 100 & 200 & 300 & 400 & 500 & 600 &  \\ \midrule
\textbf{Ridge integration method}             &     0.47 & 4.54 &  10.93& 17.63 & 23.96& 32.27  &  \\ 
\textbf{Benchmark - unconstrained}           & 0.38 &  1.79 &  5.38 & 11.79 & 17.95 & 23.08 &  \\
\textbf{Benchmark - frozen}           & 1.00   &  8.33 & 10.00 & 22.67 & 20.00 & 43.00 &   \\ \hline
\multicolumn{1}{c}{N$_2$ - Temperature (K)} & 100 & 200 & 300 & 400 & 500 & 600 &  \\ \midrule
\textbf{Ridge integration method}       & 8.75 & 12.89 & 17.92 & 26.10 & 31.43 & 41.26 &     \\
\textbf{Benchmark - unconstrained}           &  5.00  &  15.31 & 23.44 & 18.12 &  34.37 &  34.69 &   \\
\textbf{Benchmark - frozen}            & 6.97 & 13.94 &  14.79 & 21.63 & 35.02 & 38.22 &     \\ \bottomrule
\multicolumn{1}{c}{CO$_2$ - Temperature (K)} & 100 & 200 & 300 & 400 & 500 & 600 &  \\ \midrule
\textbf{Ridge integration method}        & 283.72 & 29.38 & 17.16 & 13.97 & 14.01 & 15.59 &    \\
\textbf{Benchmark - unconstrained}         &  91.85 & 26.3 & 18.89 & 11.94 & 13.33 & 14.72 &   \\
\textbf{Benchmark - frozen}         &  107.04 & 28.15 & 16.67 & 13.89 & 12.22 & 10.83 &   \\ \bottomrule
\end{tabular}
\end{table}

\subsection{Computational cost and accuracy} \label{sec:cost_acc}
\label{sec:con}
\begin{figure}[H]
     \centering
     \begin{subfigure}[b]{0.5\textwidth}
         \centering
         \includegraphics[width = \textwidth]{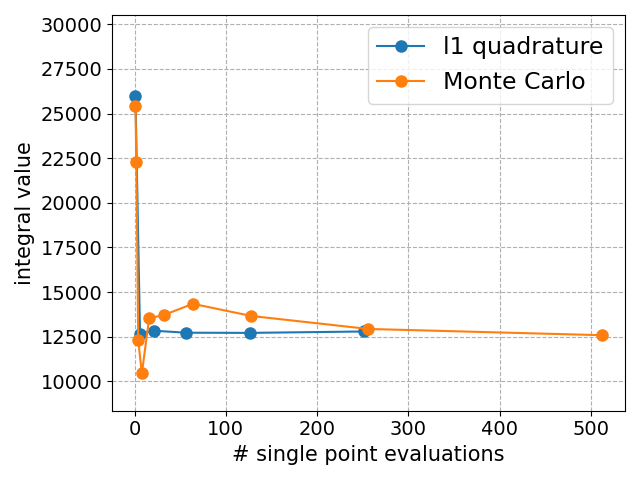}
        \caption{ridge integral of N$_2$ (Equation \ref{RIDGERR})}
        \label{fig:k_Conv}
     \end{subfigure}%
     \begin{subfigure}[b]{0.5\textwidth}
         \centering
        \includegraphics[width = \textwidth]{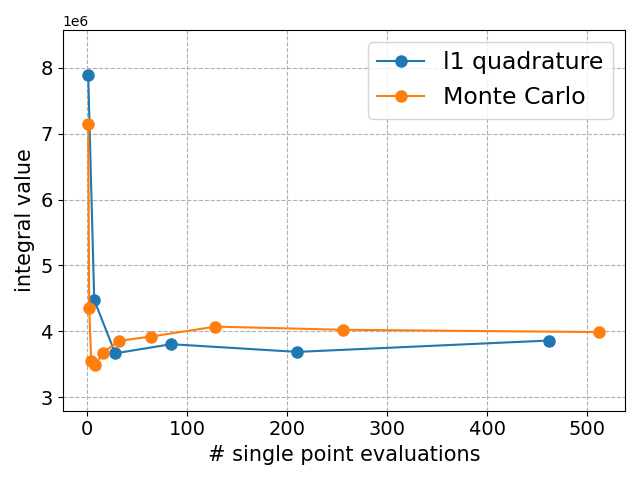}
    \caption{partition sum N$_2$ (Equation \ref{ZRT})}
    \label{fig:Z_Conv}
     \end{subfigure}
        \caption{Convergence properties of the Monte Carlo importance sampling and the $l$1 quadrature technique, tested for the required integrals (left: ridge integral, right: partition sum) in the case of N$_2$ at 300~K. Integral values are plotted as a function of the number of single point calculations.}
        \label{fig:three graphs}
\end{figure}

We begin this section by testing the two integration routines we proposed to evaluate the partition sum and the ridge integral (Equations \ref{ZRT} and \ref{RIDGERR} respectively). To give an example, we show, for the case of N$_2$, the convergence properties of the integrals of interest as a function of the number of single point evaluations (Figure \ref{fig:Z_Conv}), for both integration methods implemented: Monte Carlo importance sampling (see Section \ref{sec:l1}) and the $l$1-quadrature (Section \ref{sec:mcis}). 

Figure \ref{fig:Z_Conv} shows that for the exemplary case of N$_2$ an uncertainty of 10\,\% is reached after well below $100$ single point evaluations for the ridge integral and the total partition sum \--- a dramatic reduction of the computational expense compared to the MD benchmark. The Monte Carlo method reaches the 10\,\% uncertainty later but still around a value of 100 evaluations. This behavior could be observed for all three molecules, which makes the l1 quadrature our preferred method. Averaging over all three examined molecules (T = 300~K) one finds that the computational effort required to obtain the reaction rate (partition sum + ridge integral) using the l1 quadrature, lies around $100$~point evaluations or below. 
Next we want to compare this result to the more established methods (see Table~\ref{tab:accu}). For Eyring and ridge based predictions, we define the relative accuracy as the averaged relative deviation from the MD benchmark (unconstrained case). The averaging is done over all temperatures and molecules. For example, a relative accuracy of two means that, on average, a given method will deviate from the benchmark by a factor of two. For the MD-simulation we employ probability theory to determine the standard deviation of the number of counted transition events. Estimates for the expected numbers of single point evaluations are chosen such that the given relative accuracy value is reached.

\begin{table}[H]
\caption{Overview of accuracy and number of required single point calculation to reach the given accuracy for all methods discussed. See text for details.} \label{tab:accu}
\centering 

\begin{tabular}{lcccccl}
        \toprule
        \textbf{method}    & \textbf{rel. acc.}  & \textbf{\# of points} & \textbf{evaluated via}             \\ 
        \hline
        MD (unconstr.) &  1.1     & $10^7$     & std. deviation         \\
        Ridge method & 1.5     & $~100$    &  av. rel. deviation \\
        Eyring & 15.4 & $\approx$ 1000      & av. rel. deviation  \\
        Eyring (corr.) & 18.8 & $\approx$ 1000      & av. rel. deviation  
        \\ \bottomrule
    \end{tabular}
\end{table}

\begin{table}[H]
\caption{Overview of computational costs for all methods discussed. We show the order of core seconds consumed by the force field and DFT methods; see text for details. The value given for the ridge integration method does not include the time required for numerical integration ($\approx 10^2$ core seconds).} \label{tab:time}
\centering 
\begin{tabular}{lcccccl}
        \toprule
        \textbf{method}    & \textbf{core-seconds (FF)}  & \textbf{core-seconds (DFT)}             \\ 
        \hline
        MD (unconstr.) &  $\approx 10^5$     & $\approx 10^{11}$     \\        Ridge method & $\approx 1$     & $\approx 10^6$      \\
        Eyring &$\approx 10$ & $\approx 10^7$        \\
        Eyring (corr.) & $\approx 10$ & $\approx 10^7$      
        \\ \bottomrule
    \end{tabular}
\end{table}

An Eyring calculation can be expected to require about $1000$ evaluations due to the costly procedure of saddle point search and determination of Hessians at minimum and TS. Hence, using the $l$1 quadrature instead, with only $100$ required evaluations, the ridge method is clearly outperforming the Eyring methods in terms of computational cost, with a relative accuracy of $1.5$. Table \ref{tab:accu} clearly shows that the Ridge method produces results that are an order of magnitude more accurate than both Eyring-based predictions. The MD-simulation of course outperforms the ridge method, but only at the cost of an unfeasible number of single point evaluations. As can be seen in Table~\ref{tab:time}, employing DFT-based MD-simulations would require approximately $10^8$ core hours, which is currently only feasible for larger supercomputer facilities. Whether it is actually necessary to resort to energy predictors at the level of DFT is a separate question, which will be addressed in Section~\ref{sec:Methval}.

\subsection{Comparison of force-field, tight-binding and DFT calculators}
\label{sec:Methval}
In Section \ref{sec:con} we could clearly demonstrate the superiority of the ridge protocol over the Eyring-based methods. In a next step, we investigate the performance of the ridge method at a level of theory where full Born-Oppenheimer Molecular Dynamics is no longer an option. We start with the examination of numerical values obtained for the ridge integral, the total partition sum, and the respective reaction rates for the main constituents of natural gas or bio-gas, N$_2$, CH$_4$ and CO$_2$ for a temperature of 300~K according to the methods developed in Section \ref{sec:Methods}. 
The energy predictors used are the GFN force field \cite{grimme2017robust}, the GFN2-xTB tight binding ansatz,\cite{Bannwarth2019,Bannwarth2021} a rung two DFT functional (B97-D\cite{Grimme2006} with basis cc-pVDZ\cite{Dunning1989}) and a rung four DFT functional ($\omega$B97x-D3\cite{Lin_Chai} with basis cc-pVTZ), in hierarchical order. The DFT calculations were carried out with the Q-Chem computational chemistry package.\cite{qchem-short}

\begin{figure}[H]
     \centering
     \begin{subfigure}[b]{0.33\textwidth}
        \centering
        \includegraphics[width=\textwidth]{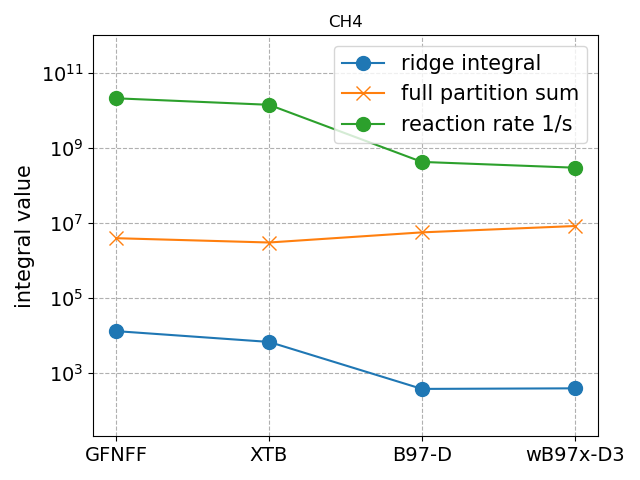}
        \caption{Methane}\label{fig:comp-CH4}
     \end{subfigure}%
     \begin{subfigure}[b]{0.33\textwidth}
        \centering
        \includegraphics[width=\textwidth]{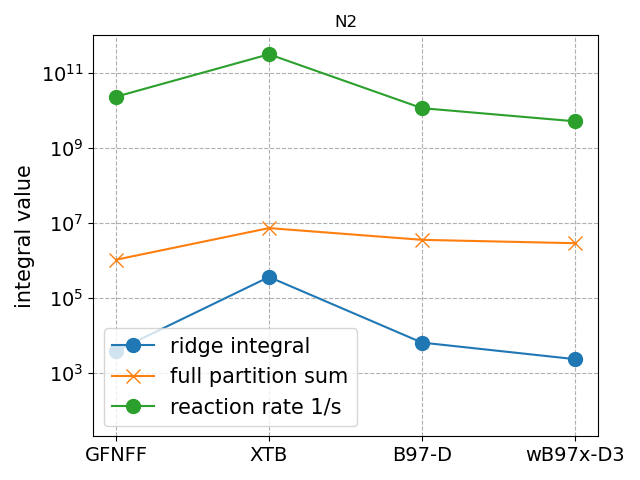}
        \caption{Molecular nitrogen}\label{comp-N2}
     \end{subfigure}%
     \begin{subfigure}[b]{0.33\textwidth}
        \centering
        \includegraphics[width=\textwidth]{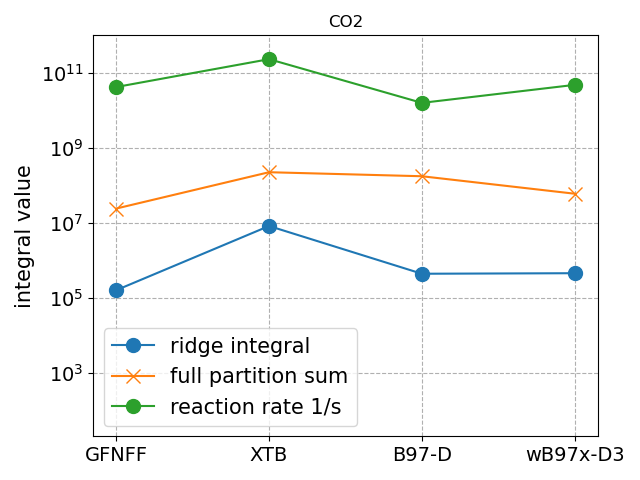}
        \caption{Carbon-dioxide}\label{fig:comp-CO2}
     \end{subfigure}
        \caption{Comparison of different levels of energy predictors. For each gas species, the partition sum (Equation \ref{ZRT}), the ridge integral and the reaction rate (Equation \ref{P_trans}) are plotted as a function of the underlying energy predictor. The integration of the partition sum is carried out over the volume defined by the constraining sphere introduced in Section~\ref{sec:Ass_ridge}.}\label{fig:levels of theory}
\end{figure}

Our results are summarized in Figure~\ref{fig:levels of theory}. In all cases, significant deviations between the partition sums and ridge integrals calculated with the lower level methods (GFN-FF and GFN2-xTB) and the DFT based predictions are apparent. These deviations get as big as two orders of magnitude. For the other examined molecule CH$_4$ the individual value of the ridge integral as well as the reaction rate deviate by two orders of magnitude. This fact suggests that the accuracy provided by the examined force fields is not nearly sufficient to describe common pore propagation problems. As expected, the DFT based methods are significantly more reliable. With one exception (N$_2$), for all examined quantities and molecules, the results provided by the rung two and rung four functionals differ by less than a factor of three. Note that differences due to the choice of functional are therefore even larger than the methodological error of the ridge method in comparison to the MD reference (compare with Figure~\ref{fig:reac-rate-comp}, transitions per ns). Another noteworthy examination is that the usage of GFN2-xTB did only marginally improve the results of the studied quantities compared to the GFN force field (compared to the DFT benchmark results). 

\subsection{Realistic selectivities for gas separation}
\label{sec:gassep}
\begin{figure}[H]
	\centering
		\includegraphics[scale=.20]{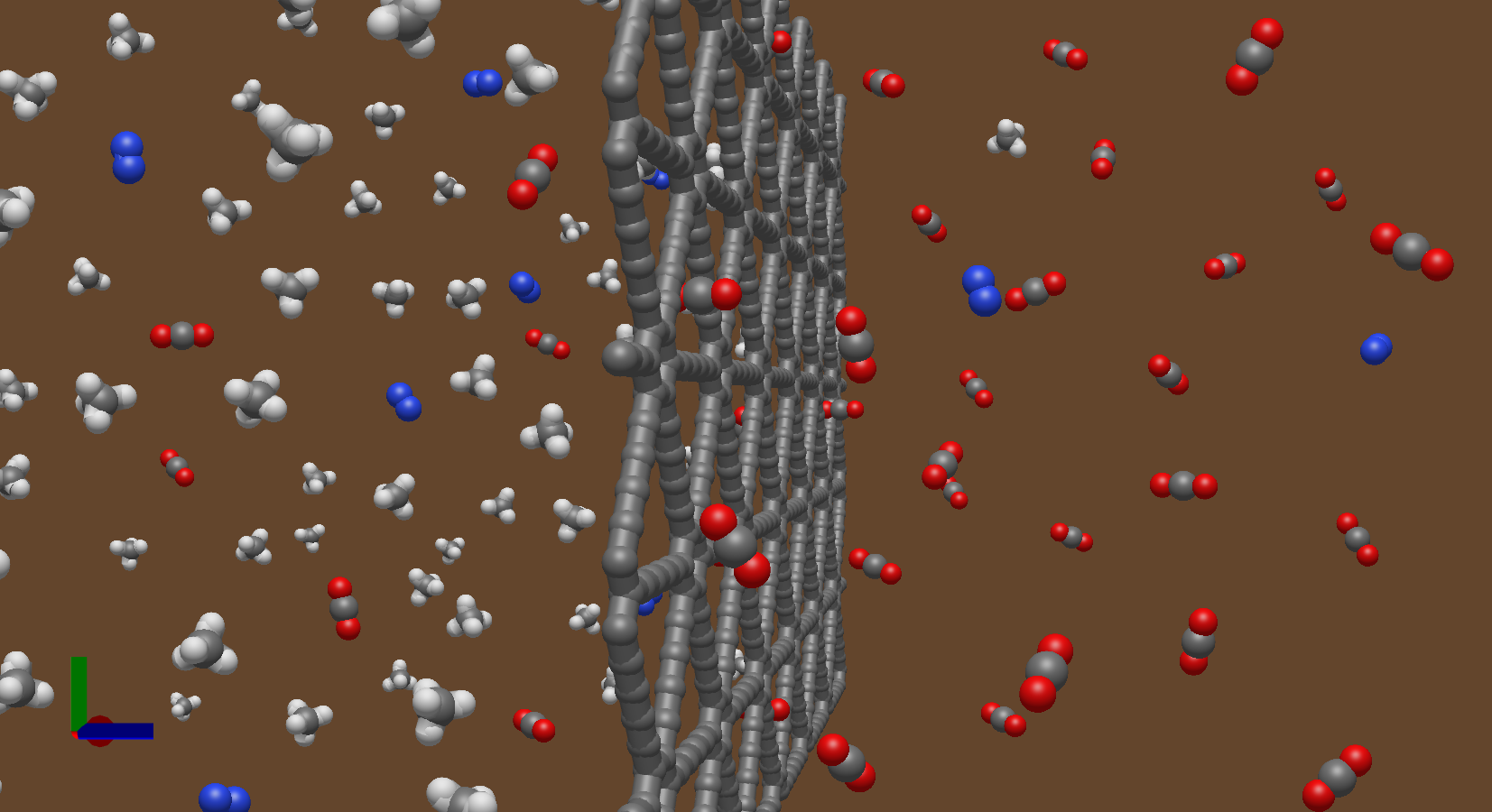}
	  \caption{Graphical illustration of a proposed mechanism for natural gas purification. The left chamber is contracted and unwanted molecules are hoped to permeate from the left chamber to the empty chamber on the opposing side. This graphic was generated using the Avogadro software package\cite{avogadro}}
	  \label{fig:piston}
\end{figure}

A last step, after having shown the correctness and the effectiveness of the ridge method, is to derive macroscopic parameters for a realistic gas sieving scenario involving the molecules CH$_4$, N$_2$ and CO$_2$ and a single-layered graphdiyne membrane. In a thought experiment, the latter may separate a given volume into two chambers of equivalent size as it is illustrated in Figure \ref{fig:piston}. A pressure of $1$ ATM and a temperature of 300~K is assumed on both sides. We further assume a perfect, periodically extended membrane built from identical model pores as described in Section~\ref{sec:Methods}, and neglect any inter-molecular interactions of the gas phase molecules. 

The latter simplification requires justification since pore clogging through adsorption might be a relevant feature of the porous membrane. A possible check is to find the fraction of occupied adsorption minima at the given temperature and pressure conditions. We define a adsorption minimum to be occupied if a given molecule of species $i$, is located within a sphere centered at the pore center with a radius of $3$ angstrom as illustrated in Figure~\ref{fig:TRANS_PORE}. 
The associated average occupation number is obtained by multiplying the probability that one given molecule of species $i$ occupies the adsorption minimum by the total number of molecules of that species $N_i$ in
the chamber. The adsorption probability is given by the ratio of the partition sums of the molecule
constrained to the sphere $Z_{\rm sphere}$ and the whole piston chamber $Z_{chamber}$. Hence,
\begin{equation}
    \langle N_{i,occ} \rangle = \frac{Z_{\rm sphere}}{Z_{\rm chamber}} N_i,
    \label{eqn:occprob}
\end{equation}
where the associated average occupation number is denoted as $\langle N_{i,occ} \rangle$.
If the average occupation number of all species is small it is sensible to expect that the pore transition of a molecule is very rarely hindered by another molecule.

For a pressure of $1$ ATM and a temperature of 300~K we find values $\langle N_{i,occ} \rangle \approx 0.01$, with $N_i$ derived via the ideal gas law. Therefore, propagation hindrance is highly unlikely, and it is sensible to consider the molecules as non-interacting in terms of pore-clogging. With that, we can identify the total flow rate of species $i$ through a single pore as the product of the probability $p_{i}$ that a certain molecule propagates through the pore and the number of molecules $N_i$ in that chamber. Probability $p_{i}$ can be obtained via Equation \ref{RIDGERR}, where we integrate over a small region in the vicinity of the pore center. To find the denominator of Equation \ref{RIDGERR}, the total partition sum, the integration domain $V$ of Equation~\ref{ZRT} is now the whole chamber volume. Multiplying our result with the number of pores $g$ per unit area and the number of molecules $N_i$ of species $i$ within the chamber with volume $V$ at a pressure of 1~Pascal we obtain the specific molar flow rates $Q_i$ of the molecules of interest via
\begin{equation}
    Q_i = \frac{p_{i} g N_i}{N_A}.
    \label{eqn:flowrate}
\end{equation}
Consequently, $Q_i$ is given in units of $mol $\space$ m^{-2} s^{-1} pa^{-1}$, with $N_A$ denoting the Avogadro constant. Note that the flow rate does not depend on the volume $V$, since the volume appears linearly in the total partition sum for $p_i$ and in the number of molecules in the chamber if we assume an ideal gas in both cases. 

\begin{figure}[H]
     \centering
     \begin{subfigure}[b]{0.33\textwidth}
         \centering
         \includegraphics[width = \textwidth]{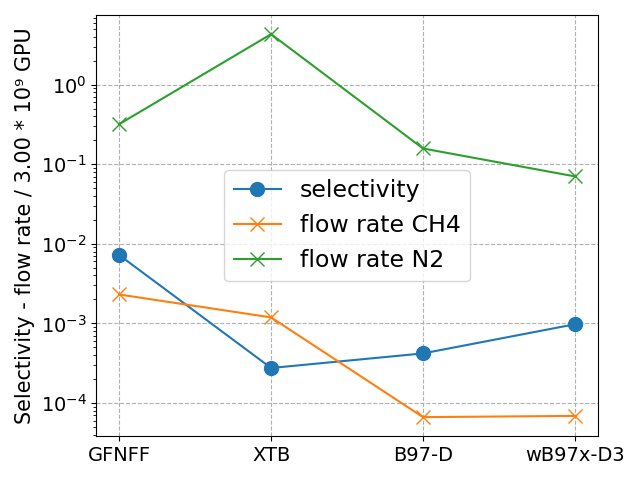}
        \caption{CH$_4$/N$_2$ }
        \label{fig:selec1}
     \end{subfigure}%
     \begin{subfigure}[b]{0.33\textwidth}
         \centering
         \includegraphics[width =\textwidth]{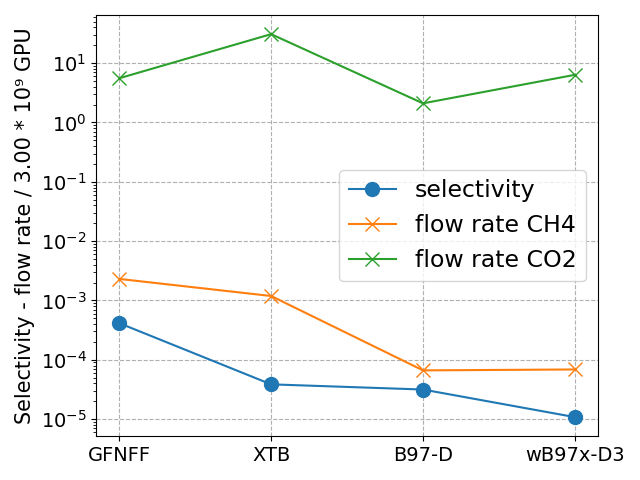}
        \caption{CH$_4$/CO$_2$}
        \label{fig:selec2}
     \end{subfigure}%
     \begin{subfigure}[b]{0.33\textwidth}
         \centering
         \includegraphics[width = \textwidth]{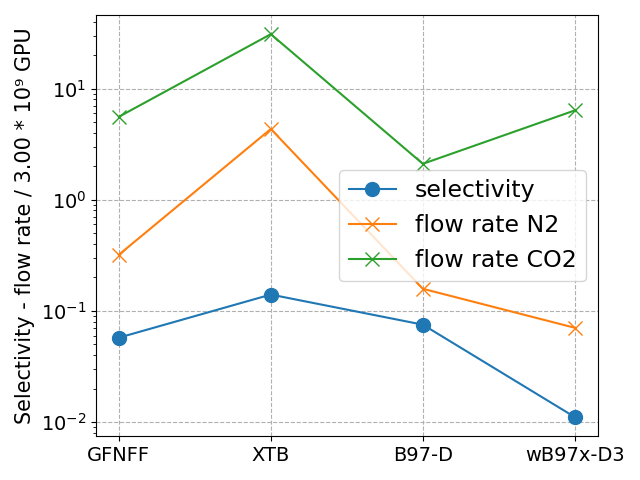}
        \caption{N$_2$/CO$_2$}
        \label{fig:selec3}
     \end{subfigure}
        \caption{The plot features the selectivity as well as the flow rates calculated via the ridge method. The flow rates are given in units of $10^9 \times 3$ GPU which is equivalent to 1 $mol $\space$ m^{-2} s^{-1} pa^{-1}$.}
        \label{fig:selectivity}
\end{figure}

The flow rates, as well as their ratio, i.e. the selectivity, are presented in Figure \ref{fig:selectivity} for all four energy predictors studied in this work. The selectivities calculated via the DFT methods differ by a factor of two for CH$_4$/N$_2$ and CH$_4$/CO$_2$ and about a factor of three for N$_2$/CO$_2$. From this we can deduce that, employing DFT methods, the order of magnitude of the obtained selectivity results should be correct. For the GFN force field, the picture is different. The CH$_4$/N$_2$ selectivity differs by more than two orders of magnitude from the DFT result. On first sight, the GFN2-xTB calculator delivers better results. However, for this calculator, as well as for GFN-FF, the errors in the obtained flow rates can be as high as two orders of magnitude. Hence, error cancellation in favor of the GFN2-xTB predictor is apparent. 
The fact that CH$_4$/N$_2$ and CH$_4$/CO$_2$ selectivities are far below one even at DFT level suggests that graphdiyne is suitable for the task of purifying natural gas. 

Using the obtained flow rates at DFT level (wB97x-D3 with basis cc-pVTZ) we can calculate the composition of the total flow through the membrane. Assuming that natural gas is composed of 94\,\% CH$_4$ (or other relevant carbon composites), 5\,\% N$_2$ and 1\,\% CO$_2$ \cite{naturalgas1} we obtain a composition of the total flow of 94.7\,\% CO$_2$, 5.2\,\% N$_2$ and 0.1\,\% CH$_4$. The high flow rates of CO$_2$ suggest that graphdiyne might be a suitable membrane for CO$_2$ removal.

\section{Discussion}
The ridge integration method provides excellent predictions for the propagation rates of N$_2$, CH$_4$ and CO$_2$ through graphdiyne. Moreover, a tremendous advantage can be seen in terms of computational effort compared to a molecular dynamics simulation. In our case, the latter requires $10^6$ to $10^7$ single-point PES evaluations. In general, the number of evaluations needed to obtain reasonable estimates through direct sampling can become astronomically large, depending on the actual height of the barrier. The ridge integration method, on the other hand, yields reasonable estimates with about $100$ evaluations of the PES, at any barrier height, providing enormous savings in computing time. It is hence paving the way to employing more costly energy predictive methods such as density functional theory. We note that, by cutting the cost to this level, the ridge integration method requires less PES evaluations as the sampling of a single trajectory would typically necessitate in order to connect the minima on either side of the pore. This fact is ruling out the superiority of all trajectory based importance sampling methods in terms of computational cost.

In a direct comparison of all three studied approaches, full MD simulation, Eyring theory and ridge integration, all of them utilizing the same method for single point evaluation, the MD simulation is obviously delivering the most precise result. The precision of Molecular Dynamics simulations, however, comes at the price of an unfeasible number of single point evaluations, typically only manageable for low-level calculators such as force field approximations. Unfortunately, as we have shown in Section~\ref{sec:Methval} the energy values provided by these low-level calculators can falsify the results by more than two orders of magnitude. Other studies confirm these findings, but suggest even higher potential errors for force field-based investigations.\cite{fruehwirth2016chiral} Eyring theory, on the other hand, needs PES evaluations at the TS and the minima only, together with their second derivatives, and preliminary evaluations for the localization of the actual extreme points on the PES. Although the latter can become costly if Hessians have to be evaluated numerically, and the extreme points need to be found in the first place, this makes Eyring theory a suitable tool also for more demanding PES calculators. However, for the problem at hand, the method turns out to be intrinsically imprecise, as it produces deviations of more than an order of magnitude for the reaction rate on average, which can potentially lead to selectivity prediction errors in the range of two orders of magnitude. \emph{Ad hoc} corrections of the partition sum contributions can improve these results in some cases, but can also worsen them in others. These results clearly leave the ridge method as the only viable option. It provides values closest to the MD benchmark at the lowest number of PES evaluations needed and delivers excellent predictions for the CH$_4$, CO$_2$ and N$_2$ reaction rates. Moreover, it demands the lowest computational cost of all examined methods, thereby enabling usage of high level energy calculators such as DFT. 
However, our current implementation of the ridge ansatz shares with Eyring Theory the inability to include inter-molecular interactions. Molecular Dynamics simulations offer the straightforward inclusion of these effects, but this advantage comes again at the price of an even higher number of single point evaluations. Fortunately, as was shown in Section \ref{sec:gassep}, inter-molecular interactions, anticipated in the form of pore clogging, can be neglected for natural gas purification at ambient pressure. Hence, it can be expected that the ridge method reproduces the hypothetical result of an DFT-based MD-simulation of membrane-based molecular sieving scenarios with a relative accuracy of 1.5 as shown in Table \ref{tab:accu}.

The suitability of monoatomic layers for the purification of natural gas or CO$_2$/N$_2$ separation for carbondioxide storage has been assessed in the past. \cite{schrier2012carbon}, \cite{zhao2017promising}
In the work by Schrier, a two-dimensional hydrocarbon polymer, PG-ES1, was studied via MD simulations yielding a CO$_2$/N$_2$ selectivity around 100, which is similar to our ridge-based predictions for graphdiyne obtained at DFT level.\cite{schrier2012carbon} However, the big advantage of graphdiyne is the significantly higher flow rates for all studied molecules due to the larger pore size (For CO$_2$ the graphdiyne values would be $\approx 10^{10}$ compared to $\approx 10^5$ GPU for PG-ES1).
Zhao and coworkers\cite{zhao2017promising} investigated three graphdiyne-like membranes, which were designed by substituting one-third diacetylenic linkages with heteroatoms hydrogen, fluorine, and oxygen (GDY$_X$, X = H, F, and O), respectively. Comparing our results to this study we find similar selectivities but higher gas permeation rates (by two to three orders of magnitude). In the case of unmodified graphdiyne, favourable selectivities, in combination with the high gas permeation rates, suggest a better suitability for natural gas purification or CO$_2$ filtering at an industrial scale.

\section{Conclusion and outlook}
We presented an alternative method for the calculation of propagation probabilities as required in studies of molecular sieving and molecular diffusion. Given the example of CH$_4$ separation from N$_2$ and CO$_2$ via graphdiyne, we show that Eyring theory, the starting point of our method development, provides imprecise transition probabilities for this type of problem, which leads to incorrect predictions regarding membrane permeance and gas selectivity. Pronounced anharmonicities and low-lying vibrational frequencies, inherent features of molecular sieving scenarios, are not covered in the original theory.

Taking advantage of the geometry of the typical problem setting, we introduce the `ridge', a hypersurface which divides the reaction volume into a reactant and a product side. We use this concept to calculate accurate partition sum contributions through combination of Monte Carlo importance sampling and l$1$ quadrature. For the separation of methane from carbon dioxide and nitrogen via graphdiyne we show that the transition probabilities based on Eyring theory are off by about one order of magnitude, while the ridge integration method is able to reproduce the benchmark result obtained from molecular dynamics simulations over a temperature range of 500~K within 50\,\%. This substantial improvement can be achieved with approximately $100$ single point evaluations, using the l1-quadrature for the necessary integration tasks. This number of external PES evaluations is even smaller than what is required by a typical Eyring calculation and hence orders of magnitude smaller than what is needed to obtain reasonable statistics from MD simulations. This feature makes our method a suitable tool for future treatments of pore propagation events via more accurate and therefore more expensive methods such as density functional theory. Furthermore, we showed that it is clearly necessary to employ these higher level energy predictors as the results provided by force fields may deviate from correct values by one or two orders of magnitude. Finally, using DFT as a high-level predictor and employing the ridge method, we could show that graphdiyne is a suitable material for natural gas or bio-gas purification. The material does not significantly outperform other studied membranes in terms of selectivity but yields higher gas permeation rates.

We emphasize that the problem of pore propagation has been chosen to showcase the capabilities of our method, but we expect that it can easily be generalized to more complex scenarios. Applications can be anticipated for problems of molecular diffusion within periodic structures such as zeolites or metal-organic frameworks for gas separation and storage. These processes can be interpreted as a hopping between chambers \cite{keil2000modeling} separated by clearly defined ridges. 
Future improvements of the ridge method might involve the inclusion of inter-molecular interactions, solvent effects, quantum corrections to the relevant partition sum factors, and the removal of stiffness constraints, in particular when describing larger biomolecules, e.g. for membrane-based chiral separation of racemic mixtures.

\begin{appendix}
\section{Crossing volume}
\label{A1}
To understand which points in phase space lie within the earlier defined `crossing volume' it is beneficial to look at a graphical illustration of the problem in 2D:

\begin{figure}[H]
    \centering
    \includegraphics[width = 0.45\textwidth]{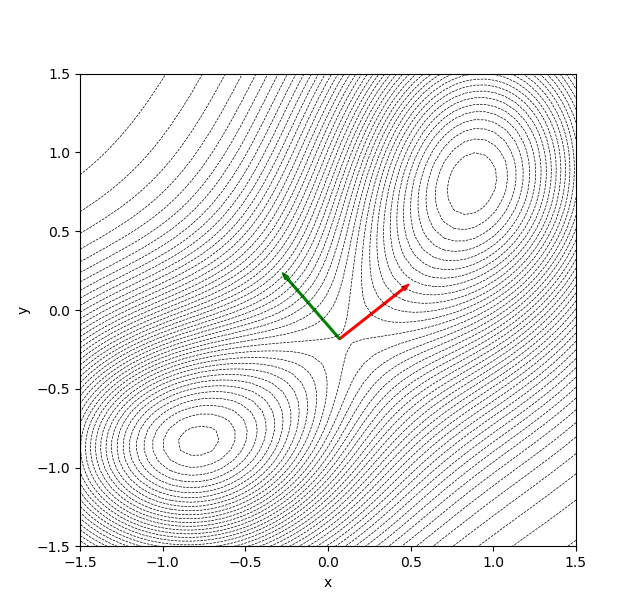}
    \caption{2D PES with two minima and one saddle point. The red arrow points in the direction of the reaction coordinate, the green arrow in the direction normal to the latter.}
    \label{fig:1.1}
\end{figure}

The separating hypersurface, which is just a line in 2D, must be chosen such that it is normal to the reaction coordinate (red arrow). It is hence represented by a line parallel to the green arrow.

In accordance with Eyring theory we assume that trajectories that cross the barrier far away from the saddle point are very rare due to their small Boltzmann weight. Therefore, it is justified to perform a second order Taylor expansion at the saddle point,

\begin{equation}
    \tilde{V}(x,y) \approx \frac{k_{x}}{2}x^{2} + \frac{k_{y}}{2}y^{2},
    \label{eqn:2.6}
\end{equation}
with the coordinate origin set to the saddle point and the $x$ and $y$ axes pointing along the colored arrows (eigenvectors of the Hessian).

To determine the crossing volume, we  note that there should be no constrains, neither on the $x$-velocity, nor on the $x$ spacial coordinates.
Furthermore, in order to calculate the transition probability from $-y$ to $+y$, we know that $v_y$ has to be positive. Moreover, since a trajectory needs sufficient velocity to cross within the time interval $\delta t$, the upper limit of $y$ is zero and the lower limit is $-v_{y} \delta t$.
Having determined the boundaries of the hypervolume, we can write down an expression for the probability of finding the particle inside that volume by using Equation \ref{prob_Phase}:

\begin{equation}
    P_{\text{trans}}(\delta t) = \frac{1}{Z}\int_{-\infty}^{\infty}\int_{-v_{y} \delta t}^{0} e^{-\beta \tilde{V}(x,y)} \, dx\, dy \int_{-\infty}^{\infty}\int_{0}^{\infty} e^{-\beta{\frac{p_{x}^2 + p_{y}^2}{2m}}} dp_{x}\,dp_{y}.
\end{equation}

In the limit of $\delta{}t\rightarrow{}0$ the integral over $dy$ can be  evaluated instantly. It is just the integrand at $y=0$ multiplied by $v_{y} \delta t$. With $v = \frac{p}{m}$ we obtain

\begin{equation}
    P_{\text{trans}}(\delta t) = \frac{\delta t}{Z}\int_{-\infty}^{\infty} e^{-\beta \tilde{V(x,0)}} \, dx \int_{-\infty}^{\infty}\int_{0}^{\infty} e^{-\beta{\frac{p_{x}^2 + p_{y}^2}{2m}}} \frac{p_{y}}{m}  dp_{x}\,dp_{y}.
    \label{eqn:2.8}
\end{equation}
Using the approximated potential given by Equation \ref{eqn:2.6} we obtain

\begin{equation}
    P_{\text{trans}}(\delta t) = e^{-\beta \Delta E} \frac{\delta t }{Z}\int_{-\infty}^{\infty} e^{-\beta \frac{k_{x}}{2}x^{2}} \, dx  \int_{-\infty}^{\infty}\int_{0}^{\infty} e^{-\beta{\frac{p_{x}^2 + p_y^2}{2m}}} \frac{p_{y}}{m}   dp_{x}\,dp_{y}.
    \label{standardint}
\end{equation}
The zero point of the potential shall be set to the energy of the minimum and $\Delta E$ can be pulled out, with $\Delta E$ being the energy difference between the minimum and the transition state.
The RHS of Equation \ref{standardint} is a product of two Gaussians and a standard integral. These can all be easily evaluated, yielding

\begin{equation}
    P_{\text{trans}}(\delta t) = e^{-\beta \Delta E} \frac{\delta t}{Z} m^{\frac{1}{2}} \frac{2\pi}{ \beta^{2} \sqrt{k_{x}}}.
    \label{eqn:2.10}
\end{equation}
$Z$ can be approximated by a similar quadratic expansion around the minimum and is also just another product of four Gaussians:

\begin{equation}
    Z = \int_{-\infty}^{\infty} \int_{-\infty}^{\infty} e^{-\beta V(x,y)} \, dx dy \int_{-\infty}^{\infty}\int_{-\infty}^{\infty} e^{-\beta{\frac{p_{x}^2 + p_{y}^2}{2m}}}   dp_{x}\,dp_{y} = \frac{4\pi^2 m}{\beta^2 \sqrt{k_{x,min} k_{y,min}}}.
    \label{eqn:2.11}
\end{equation}
With Equations \ref{standardint}, \ref{eqn:2.10} and the initial definition of the reaction constant 
$k=\frac{P(\delta t)}{\delta t}$ we obtain

\begin{equation}
    k = e^{-\beta \Delta E} \frac{\sqrt{k_{x,min} k_{y,min}}}{2 \pi \sqrt{k_{x,TS} m}}
    \label{eqn:2.12}.
\end{equation}

\section{Simplifications regarding the mobile phase}
\label{A2}
Our treatment of the mobile phase is based on two assumptions. First, the potential $V$ can be divided into an vibrational and a rotational-translational part, since $V_1$ depends only weakly on $t_i$ and $r_i$.

\begin{equation}
    V(\xi_1,\xi_2,...,\xi_n) = V_1(v_1,v_2,...,v_{n-6}) +  V_2(t_1,t_2,t_3,r_1,r_2,r_3).
    \label{int}
\end{equation}

The latter is justified because the intramolecular interaction producing $V_1$ is orders of magnitude stronger than the van der Waals interaction leading to $V_2$. Substituting the potential defined in Equation \ref{int} into Equation \ref{RIDGERR}, we find that the integrals over $v_i$ as well as the integrals over all $p_i$ with $i \neq 3$ are appearing in both the numerator and in Z. Thus, they cancel, leading to Equation \ref{P_trans}.

Second, we neglect the dependence of $\boldsymbol{\tau_i}$ and $\boldsymbol{\rho_i}$ on $v_i$. This is justified since we expect only very small changes in the vibrational degrees of motion. Furthermore, we use the fact that $\boldsymbol{\nu}$ are normalized and orthogonal to $\boldsymbol{\tau_i}$ and $\boldsymbol{\rho_i}$. Thus, we can effectively neglect the dependence on vibrational coordinates in $S_{\text{TRV}}$ and $J_{\text{TRV}}$ from Equations \ref{eqn:STRV} and \ref{eqn:JTRV}.

\section{$l$1-quadrature}
\label{A:l1_quadrature}
For a given (non-negative) integration measure $\omega(\boldsymbol{x})$ and for a set of given abscissas $\{\boldsymbol{x}_i\}_{i=1}^N$ we want to determine the weights $\{w_i\}_{i=1}^N$ such that a certain set basis functions $\{b_\alpha(\boldsymbol{x})\}_{\alpha=1}^n$, that contains the constant function, is integrated exactly, whilst a minimal number of weights should be unequal zero, i.e. a minimal number of the proposed abscissas is used. The integration formula reads
\begin{equation}
    I(f) := \int_{\Omega} f(\boldsymbol{x}) \omega(\boldsymbol{x}) \;\mathrm{d} x \approx \sum_{i=1}^N w_i f(\boldsymbol{x}_i) =: Q(f),
    \label{quadrature_weight}
\end{equation}
for an arbitrary function $f$. Since the numerical integration should be exact for all basis functions we get the condition
\begin{equation}
    I(b_\alpha) \overset{!}{=} Q(b_\alpha) = \sum_{i=1}^N w_i b_\alpha(\boldsymbol{x}_i) \qquad \forall \alpha\in\{1,\dots,n\},
    \label{eq:basis_exact}
\end{equation}
which can be rewritten as a linear equation $M \boldsymbol{w} = \boldsymbol{r}$, where $M_{\alpha,i} = b_\alpha(\boldsymbol{x}_i)$ and $r_\alpha = I(b_\alpha)$. Note that in order to obtain $\boldsymbol{r}$ one needs to integrate the corresponding basis functions; this can be done with a high precision quadrature rule, since the integration measure is obtained via the low level potential. However, calculations showed that even convergence to a relative error of $10^{-2}$ for the right hand side $\boldsymbol{r}$ give useful results for the integral of interest.

Since the $n$ basis functions are linear independent the Matrix $M$ has rank $n$ if $N\geq n$. We want to obtain a solutions $w_i$ such that a minimal number of weights is non zero and that the weights are positive, since this leads to much more stable numerical integration \cite{l1Quad2}. However, this problem is hard to solve. For that reason we solve a slightly different problem, that can be solved easily and maintains the desired properties \cite{l1Quad,l1Quad2}. We minimize the sum of the modulus of the weights
\begin{equation}
\sum_{i=1}^N \vert w_i \vert \rightarrow \mathrm{min}
\label{eq:minimization}
\end{equation}
under the constraint given by Equation \ref{eq:basis_exact}.
Since the basis set contains the constant function the sum of all weights is fixed by the corresponding integration condition (see Equation \ref{eq:basis_exact}), which yields a positive integral value. Thus the solution to the minimization problem yields minimal negative weights. Additionally, this minimization problem can be reformulated as
\begin{equation}
    \sum_{i=1}^N w_{+,i} + w_{-,i} \rightarrow \mathrm{min}, \quad \mathrm{s.t.} \quad
    \begin{cases} \begin{bmatrix} M & -M \end{bmatrix} \begin{bmatrix} \boldsymbol{w}_+ \\ \boldsymbol{w}_- \end{bmatrix} = \boldsymbol{r} \\
    w_{+,i}, w_{-,i} \geq 0 \qquad \forall i\in\{1,\dots,N\},
    \end{cases}
    \label{eq:simplex}
\end{equation}
where $w_i = w_{+,i} - w_{-,i}$. Equation \ref{eq:simplex} is the standard form of a linear program and thus can be efficiently solved using a simplex algorithm. Applying a simplex scheme that finds basis solutions we can ensure that at most $n$ weights are non-zero, meaning that at most $n$ single point evaluations are needed in the resulting quadrature rule. It is convenient to chose a large number of proposed abscissas in order to ensure stable (non-negative) solutions for the weights. Similar schemes are described in Ref.~\citenum{l1Quad2,l1Quad} for different applications.

We use multidimensional polynomials up to a certain degree as basis functions. The resulting quadrature rules can be constructed nested, by generating the abscissas and weights for the highest degree and then using the obtained abscissas as a proposal for lower degrees. This ensures that a convergence series of the integral of interest can be obtained without additional effort.

\end{appendix}

\section*{Acknowledgements}
This research has been supported by the Austrian Science Fund (FWF) under Grant No. P 29893-N36. Further support by NAWI Graz is gratefully acknowledged.


\begin{thebibliography}{54}
\bibitem [{fn5()}]{fn5}%
  \BibitemOpen
  \href@noop {} {}\bibinfo {note} {{A Head-Gordon damping function with
  $\alpha{}=4$ and $\omega{}_{0}=100$\,cm$^{-1}$ has been employed for the
  continuous interpolation between rotational and vibrational
  contributions.}}\BibitemShut {Stop}%
\end{thebibliography}
%

\end{document}